\documentclass[11pt]{article}
\usepackage{amsbsy}
\usepackage{fullpage}
\usepackage[dvips]{epsfig}
\usepackage{latexsym}
\usepackage{psfig}
\usepackage{epsfig}
\usepackage{rotate}
\usepackage{amsmath}

\newcommand{\ep}{\epsilon}
\newcommand{\vep}{\varepsilon}

\newcommand{\ug}{{\bf g}}
\newcommand{\un}{{\bf n}}
\newcommand{\up}{\bf{p}}
\newcommand{\uq}{\bf{q}}
\newcommand{\uu}{{\bf u}}
\newcommand{\ux}{{\bf x}}
\newcommand{\TI}{{\bf T}^i}
\newcommand{\TJ}{{\bf T}^j}
\newcommand{\DI}{D^i}

\newcommand{\ts}{\mbox{\boldmath $S$}}
\newcommand{\uv}{\bf v}
\newcommand{\ee}{\mbox{\boldmath $e$}}
\newcommand{\bound}{\partial\Omega}

\newcommand{\om}{\omega}

\title
{Effective shear and extensional viscosities of
concentrated disordered suspensions of rigid particles.}  
\author{Leonid Berlyand\\
Department of Mathematics and Materials Research Institute\\  Penn State
University\\
University Park, PA 16802, USA\\
 ({\tt berlyand@math.psu.edu})\\
and\\ 
Alexander Panchenko\\
Department of
Mathematics\\
 Washington State University\\
Pullman WA 99164, USA \\
({\tt
panchenko@math.wsu.edu})}

\begin{document}
\maketitle
\begin{abstract}
We study effective shear viscosity $\mu^\star$ and  effective extensional
viscosity $\lambda^\star$ of concentrated non-colloidal suspensions of rigid spherical particles. The focus is on the spatially disordered arrays, but periodic arrays are considered as well. 
We use recently developed discrete network approximation techniques to obtain asymptotic formulas for $\mu^\star$ and $\lambda^\star$ as the typical inter-particle distance $\delta$ tends to zero, assuming that the fluid flow is governed
by Stokes equations. For disordered arrays, the volume fraction alone does not determine the effective viscosity. Use of the network approximation allows us to
study the dependence of $\mu^\star$ and $\lambda^\star$ on variable distances between neighboring particles in such arrays. 

Our analysis, carried out for a two-dimensional 
model, can be characterized as global because it goes beyond the local analysis of flow between two particles and takes into account hydrodynamical interactions in the entire particle array.
Previously, asymptotic
formulas for $\mu^\star$ and $\lambda^\star$ 
were obtained via asymptotic analysis of lubrication effects 
in a single thin gap between two closely spaced particles. The principal conclusion in the paper is that, in general, asymptotic formulas for $\mu^\star$ and $\lambda^\star$ obtained by global analysis are different from the formulas obtained
from local analysis. In particular, we show that the leading term in the asymptotics of $\mu^\star$ is of lower order than suggested by the local analysis (weak blow up), while the order of the leading term in the asymptotics of $\lambda^\star$  depends on the geometry of the particle array (either weak or strong blow up).
We obtain geometric  conditions on a random particle array under which 
the asymptotic order of $\lambda^\star$ coincides with the order of the local 
dissipation in a gap between two neighboring particles, and show that these conditions are generic. We also provide an example of a uniformly closely packed particle array 
for which  the leading term in the asymptotics of $\lambda^\star$ degenerates (weak
blow up).


\end{abstract}
\makeatletter \@addtoreset{equation}{section}
\def\theequation{\thesection.\arabic{equation}} 
\makeatother
\makeatletter 
\makeatother
\newtheorem{theorem}{Theorem}[section]
\newtheorem{algorithm}{Algorithm}[section]
\newtheorem{problem}{Problem}[section]
\newtheorem{lemma}{Lemma}[section]
\newtheorem{assumption}{Assumption}[section]
\newtheorem{property}{Property}[section]
\newtheorem{example}{Example}[section]
\newtheorem{corollary}{Corollary}[section]
\newtheorem{proposition}{Proposition}[section]
\newtheorem{definition}{Definition}[section]
\newtheorem{remark}{Remark}[section]
\section{Introduction}  
Concentrated suspensions are important in many industrial applications such as
drilling, water-coal slurries transport, food processing, cosmetics and ceramics manufacture. 
In nature, flows
of concentrated suspensions appear as mud slides, lava flows and soils
liquefied by the earthquake-induced vibrations (\cite{Carr}, \cite{Cous}, \cite{Shook}).

An asymptotic formula for the effective viscosity of a 
suspension of non-colloidal particles in a Newtonian fluid, 
derived  in \cite{FA}, is based on the local lubrication analysis of the 
energy dissipation rate in the
narrow gap between a pair of nearly touching particles. The 
distance between two neighboring particles in a periodic array is the small
parameter in the problem. For periodic arrays, this
{\it inter-particle distance} is uniquely determined by the volume fraction
of particles, so that the asymptotics of the
effective viscosity is obtained as a function of  the volume fraction $\phi$
that is close to the 
maximal packing volume fraction $\phi_{rcp}$.  the asymptotics of the effective
viscosity obtained in \cite{FA} has the form
\begin{equation}
\label{00}
A\epsilon^{-1}+O(\ln\epsilon),
\end{equation}
as $\epsilon\to 0$, where $\epsilon=1-(\phi/\phi_{rcp})^{1/3}$. 
The formulas for effective viscosity of 
periodic suspensions in the whole space ${\bf R}^3$ (without boundary) subject
to a prescribed linear flow,
obtained in \cite{Keller}, also rely on the local lubrication analysis. 
Asymptotic representations for the components of the effective viscosity tensor 
calculated by \cite{Keller} are of the form 
\begin{equation}
\label{001}
A\epsilon^{-1}+B\ln\epsilon +O(1),
\end{equation}
Recently, concentrated random suspensions were investigated numerically
by \cite{Brady} using accelerated Stokesian dynamics. It was observed
that the behaviour of the effective high frequency dynamic shear viscosity 
of disordered suspensions can be accurately
described by the asymptotic $B\ln \epsilon$, indicating degeneration of the leading term in the asymptotic expansions (\ref{001}) (weak blow up). The authors of \cite{Brady} also show that
their results are in good agreement with available
experimental data (\cite{ShP}, \cite{VW}). 
This suggests that for generic random suspensions, the
asymptotics of the effective viscosity defined by the (properly normalized) global dissipation rate cannot be identified with the local
dissipation rate in a single  gap. 

In this paper, we use the discrete network approximation proposed in \cite{BBP} 
to study the asymptotics of the shear effective viscosity $\mu^\star$ and 
the extensional effective viscosity $\lambda^\star$ corresponding
to general disordered particle arrays. For such arrays, the volume fraction
alone is not sufficient for determining the effective viscosity.
Therefore,  instead of $\epsilon$, we use the 
inter-particle distance parameter $\delta$ that controls the distances
$\delta_{ij}$ between neighboring particles.

The mathematical construction of \cite{BBP} accounts for the 
long range hydrodynamical interactions between the particles, and
provides an algorithm for calculation of the effective viscosity, which 
takes into account variable distances between neighboring particles in non-periodic arrays.
Furthermore, in \cite{BBP} it was observed that the leading term of the asymptotics may degenerate 
due to the external boundary conditions and geometry of the particle array, while
in the scalar case (\cite{BK}) the order of the leading term is the same for all
particle arrays that form a connected network. This paper is devoted to a detailed
study of this degeneration phenomenon. In particular, we clarify the issue
of weak versus strong blow up in the asymptotics of the effective viscosity.

The definition of the effective viscosity employed in this paper differs from the
definition in \cite{Keller}. Their definition is closely related
to the definition of the effective material properties of periodic media in homogenization theory
(see e.g. \cite{BLP}, \cite{JKO}), where these properties are defined
via analysis on a periodicity
cell. Then these properties 
are determined by the properties of constituents and geometry of the periodic array,
independently of the external boundary conditions and applied forces.
Such properties are usually defined in the limit when particle size tends to
zero while the number of particles approaches infinity.

Our definition is more directly linked to the viscometric measurements
that necessarily involve boundary conditions. 
We assume that particles of a fixed size are placed in a bounded domain
(apparatus), and the typical distance between the
neighboring particles approaches zero. Thus, the total volume fraction of particles approaches the maximal packing fraction. In this case the effective viscosity
will be influenced not just by the inter-particle interactions but also by the forces or velocities prescribed at the boundary of the apparatus. Since we work
with linear models, $\mu^\star$ ($\lambda^\star$) are independent of the applied
shear (extension) rate. However, we also show that the asymptotic order of the effective viscosities depends on the relation between
the orientation of the velocity prescribed at the boundary, shape of the boundary, and orientations of the line segments connecting pairs of 
neighboring particles.

We present sufficient conditions for degeneration
(weak blow up) and non-degeneration (strong blow up) of the leading term. 
We show that $\mu^\star$ always exhibits weak blow up, while local 
analysis alone predicts strong blow up. We also show that the asymptotic order
of $\lambda^\star$  depends on the geometry 
of the particle array. In the paper, we define a broad class of arrays of particles 
for which the leading order term of $\lambda^\star$ does not degenerate. 
This situation is typical in the sense that for a 
generic array $\mu^\star$ and $\lambda^\star$ have vastly different values, and  
their ratio depends on the inter-particle distance, 
which indicates possible non-Newtonian behaviour of the effective fluid
under the imposed boundary conditions. 
We also give an example of 
a closely packed particle array for which the leading term in 
the extensional viscosity degenerates.  
\section{Mathematical model}
We consider a concentrated suspension of rigid, 
non-Brownian, neutrally buoyant
particles in a viscous
incompressible Newtonian fluid. In the two-dimensional model, 
the suspension occupies  a 
domain $\Omega$ which is a square of side length two centered at the origin.
The boundary of $\Omega$ is denoted by 
$\partial
\Omega$. 
The upper and lower sides of $\partial\Omega$ are denoted by 
$\partial\Omega^+=\{{\ux}: x_2=1\}$ and $\partial\Omega^-=\{{\ux}:x_2=-1\}$, respectively. We 
also let ${\bf e}_1, {\bf e_2}$ denote the Cartesian basis
vectors parallel to the sides of $\partial\Omega$.
The particles 
$D^j$, $j=1,2, \ldots, N$ are modelled as 
rigid disks with centers ${\bf x}^j$, placed in $\Omega$. For simplicity, we consider
the monodisperse case, so that all particles have radius $a$. The part of $\Omega$
which is not occupied by particles is the fluid domain, denoted
by $\Omega_F$.

The fluid at low Reynolds number is 
governed by Stokes equations
\begin{equation}
\label{eq1}
\mu\Delta {\uv}-\nabla P=0,\;\;\;\;{\rm div}~{\uv}=0,\;\;\;{\rm in}\;\Omega_F.
\end{equation}
where $\mu$ is the fluid viscosity, $\uv$ the velocity field, and $P$ is 
the pressure.

In this paper, we consider the external 
boundary conditions of the shear and extensional types.
The shear type boundary conditions are given by
\begin{equation}
\label{shear-fbc}
{\bf v}=
\left\{
\begin{array}{cc}
\gamma{\bf e}_1\;\;\;&{\rm on}~\partial\Omega^+ ,\\
-\gamma{\bf e}_1\;\;\;&{\rm on}~\partial\Omega^-,\\
\end{array}
\right.
\end{equation}
where $\gamma$ is a constant shear rate.

In the case of extensional boundary conditions the velocity is prescribed as
\begin{equation}
\label{ext-fbc}
{\bf v}=
\left\{
\begin{array}{cc}
\vep(-{\bf e}_2+x_1{\bf e}_1)\;\;\;&{\rm on}~\partial\Omega^+ ,\\
\vep({\bf e}_2+x_1{\bf e}_1)\;\;\;&{\rm on}~\partial\Omega^-,\\
\end{array}
\right.
\end{equation}
where $\vep$ is a constant extension rate.
The two remaining (vertical) sides of $\partial\Omega$ are free surfaces, where
the zero traction condition is prescribed.

To define particle  velocities, we first recall that
a rigid body moving in the plane defined
by ${\bf e}_1, {\bf e}_2$ has a velocity vector of the form
\begin{equation}
\label{p-vel}
{\bf v}^j({\bf x})={\bf T}^j+\omega^j{\bf e}_3\times ({\bf x}-{\bf x}^j),\;\;\;{\bf x}\in D^j,
\end{equation}
where ${\bf e}_3$  is the unit vector perpendicular to the plane of motion. Therefore, if
${\bf r}=a{\bf e}_1+b{\bf e}_2$, then
${\bf e}_3\times {\bf r}=-b{\bf e}_1+a{\bf e}_2$.
Equation (\ref{p-vel}) shows that the   
velocity of a particle $D^j$, $j=1,2, \ldots, N$ is completely determined by two parameters: 
a constant translational velocity vector ${\bf T}^j$ and a scalar 
angular velocity $\omega^j$.  
Both ${\bf T}^j$ and $\omega^j$ are unknown and must be 
determined in the course of solving the problem.

Since each rigid disk is in equilibrium, the total force and torque
exerted on $D^j$ by the fluid must be zero, which provides the  
boundary conditions on the particle boundaries $\partial D^j$: 
\begin{equation}
\int_{\partial D^j} \ts \un^{j} \, ds = {\bf 0} \;\;\hbox{and}\;\;
\int_{\partial D^j} \un^{j} \times \ts \un^{(j)}\, ds = {\bf 0},
\;\hbox{for}\; j = 1,2 \ldots N,
\label{p-bc},
\end{equation}
where ${\bf n}^j$ is the exterior unit normal to $\partial D^j$, and
\begin{equation}
\label{def-stress}
\ts=2\mu\mbox{\boldmath $e$}({\bf v})-P\boldsymbol I.
\end{equation}
In (\ref{def-stress}) and throughout the paper, $\mbox{\boldmath $e$}({\bf v})$ denotes the strain rate tensor defined by
\begin{equation}
\label{strain}
\mbox{\boldmath $e$}({\bf v})=
\frac 12(\nabla
{\uv}+
\nabla {{\uv}}^T),
\end{equation}
the superscript $T$ stands for the transposed tensor, and $\boldsymbol I$ denotes the unit tensor.

Solving equation (\ref{eq1}) with the boundary
conditions (\ref{shear-fbc}) (or (\ref{ext-fbc})) and (\ref{p-bc})
is equivalent to minimizing the functional
\begin{equation}
W_{\Omega_F}(\uu) = \frac{\mu}{4} {\displaystyle
\sum_{i,j = 1}^n} {\displaystyle \int_{\Omega_F}} \left(\frac{\partial
u_i(\ux)}{\partial x_j} + \frac{\partial u_j(\ux)}{\partial x_i}
\right)^2 d\ux,
\label{primal} 
\end{equation}
over the function space ${\cal U}$ of admissible velocity fields $\uu$. 
This space is a space of vector functions in $\Omega$ satisfying
either (\ref{shear-fbc}) or (\ref{ext-fbc}), and such that 
\begin{equation}
\uu = {\displaystyle \sum_{j=1}^n} u_j \, {\bf e}_j,
~u_j \in H^1(\Omega_F),~ j = 1 \ldots n,\; \hbox{div}\, \uu = 0, \;
(\ref{p-vel})\,{\rm holds}.
\label{sp_U}
\end{equation}
Since the fluid velocity ${\bf v}$ is the minimizer of the variational principle
(\ref{primal})-(\ref{sp_U}), the energy dissipation rate $E$ in the fluid (defined
in equation (\ref{sus-dis} below)
can be written as
\begin{equation}
\label{fl-dis-var}
E=W_{\Omega_F}({\bf v})=\min_{{\bf u}\in{\cal U}} W_{\Omega_F}({\bf u}).
\end{equation}

In next section we will see that calculation of the effective viscosities 
essentially amounts to calculation of $E$.

\section{Effective shear and extensional viscosities.}
\subsection{Effective dissipation rates}
We suppose that the suspension can be modelled on a macroscale by a single phase viscous
fluid, called an {\it effective fluid}. The velocity
field of the effective fluid is denoted by ${\uv}^0$. The effective fluid
is subject to the same external boundary conditions as the flow of the
suspension. 

We assume that the effective 
stress tensor ${\ts}^0$  satisfies the constitutive equation
of the form
\begin{equation}
\label{const}
{\ts}^0=F({\ee}({\bf v}^0)), 
\end{equation}
where $F$ is a symmetric tensor function that does not depend explicitly on ${\bf x}$. In this paper, 
we do not derive or postulate the 
precise from of the constitutive law for the
effective fluid. Instead, we use the fundamental principle (going back to \cite{Einst}),
that 
viscous energy dissipation rate of the suspension must be equal to the dissipation
rate of the effective homogeneous fluid. The dissipation rates are
defined by
\begin{equation}
\label{sus-dis}
E=\int_{\Omega_F} {\ts}\cdot {\ee}({\uv}) d{\ux}= 
2\mu\int_{\Omega_F} {\ee}({\uv})\cdot {\ee}({\uv}) d{\ux}
\end{equation}   
in the suspension,
and 
\begin{equation}
\label{eff-dis}
E^0=\int_\Omega {\ts}^0 \cdot {\ee}({\uv}^0) d\ux,
\end{equation}
in the effective fluid. In the equations (\ref{sus-dis}), (\ref{eff-dis}), 
${\ts}\cdot {\ee}={\ts}_{ij}{\ee}_{ij}$ is the inner product of tensors.

For small particle volume fractions, (\cite{Einst}, \cite{Batch}), this principle was 
further combined
with the assumption that the effective fluid is Newtonian with a constant
effective viscosity. 
However, for concentrated suspensions this assumption is not 
validated by rigorous mathematical derivation or experimental data and at present 
the problem of finding the effective constitutive law for such suspensions is still 
under investigation.  Some experimental studies (e.g. \cite{ShP}, \cite{VW}) suggest 
that the effective fluid may  be non-Newtonian. 
Our calculations of the effective viscosity suggest that  
non-Newtonian behavior is possible for irregular (non-periodic or random) suspensions.

We use the rheological definitions of shear
and extensional viscosities as ratios of the corresponding components of the stress
and strain rate tensors. To calculate the asymptotics
of the two viscosities, we employ the network approximation introduced in   (\cite{BBP}).
We analyze the network functional (discrete dissipation rate) introduced in \cite{BBP}
and  show that the standard relation between two viscosities
which holds for Newtonian fluids (see \cite{Sh} ch. 9 for the 3D-case and Appendix A for 2D-case ) 
does not hold.
\subsection{Shear viscosity}
Suppose that a homogeneous effective fluid undergoes a steady shear flow
with the shear rate $\gamma$. The velocity field 
${\uv}_{\it sh}$ satisfies the shear type boundary conditions (\ref{shear-fbc}).
The effective shear viscosity is defined by (see (\ref{eo-shear}))
\begin{equation}
\label{shev1}
\mu^\star=\frac{S^0_{12}}{\gamma}=2 \frac{E^0}{\gamma^2 |\Omega|},
\end{equation}
where $S^0_{12}$ is the corresponding component of the effective stress tensor 
and $|\Omega|=\int_\Omega d\ux$.
Note that our definition of $F^0$ in (\ref{const}) implies that 
$\ts^0$ is constant when $\ee({\bf v}^0)$ is constant.
Since $E=E^0$, the equivalent definition is
\begin{equation}
\label{shev2}
\mu^\star=2 E \gamma^{-2} |\Omega|^{-1}=
4\mu \gamma^{-2} |\Omega|^{-1}
\int_{\Omega_F} {\ee}({\uv}_{\it sh})\cdot\ee(\uv_{\it sh})d \ux
\end{equation}
Thus the calculation of $\mu^\star$ amounts to evaluation of the total dissipation rate integral
\begin{equation}
\label{dirint}
E_{sh}=2\mu\int_{\Omega_F} {\ee}({\uv}_{\it sh})\cdot\ee(\uv_{\it sh}){\it d}\ux,
\end{equation}
where $\uv_{\it sh}$ solves  (\ref{eq1})-(\ref{p-bc}).

\subsection{Extensional viscosity}
A steady extensional flow of the effective fluid is characterized by
a constant extension rate $\vep$. The velocity ${\uv}^0_{\it ext}$ satisfies
the extensional boundary conditions (\ref{ext-fbc}).
The extensional viscosity (see, e.g. \cite{Sh}. ch.9) may be defined by
\begin{equation}
\label{ev}
\lambda^\star=\frac{S_{11}^0-S_{22}^0}{\vep},
\end{equation}
where $S_{11}^0, S_{22}^0$ are components of the effective stress tensor. 
Since 
$E^0=\int_\Omega {\ts}^0\cdot\ee({\uv})^0_{\it ext}d\ux= (S_{11}^0-S_{22}^0)\vep|\Omega|$, 
the effective
extensional viscosity can be defined 
in terms of the suspension dissipation rate $E$, as follows.
\begin{equation}
\label{eev1}
\lambda^\star=\frac{E}{\vep^2 |\Omega|}=
2\mu\vep^{-2}|\Omega|^{-1}\int_{\Omega_F} {\ee} ({\uv}_{\it ext}) \cdot {\ee} 
({\uv}_{\it ext}) d\ux,
\end{equation}
and  calculation of  $\lambda^\star$ again reduces to evaluation of the total dissipation 
rate (\ref{dirint}) with  ${\uv}_{\it sh}$ replaced by ${\uv}_{\it ext}$. 
In the remaining part of the paper we derive asymptotic formulas for the total 
dissipation rate under boundary conditions (\ref{shear-fbc}) and (\ref{ext-fbc}).

\section{The Network}
Let us consider an arbitrary distribution of circular particles (disks) $D^i$, whose centers
are points ${\bf x}^i$ in $\Omega$, for $i = 1,2, \ldots, N$. We suppose that $N$
is close to $N_{\hbox{max}}$, so that neighboring particles can be close to
touching one another. 
The network consists of vertices ${\bf x}^i$ and edges. The edges connect only 
vertices that correspond to 
`` neighboring ``  particles. Note that while for a periodic array  
the notion of a neighboring vertex (particle) is obvious, for 
non-periodic (e.g. random) arrays of particles it is not immediate. We introduce it
via a Voronoi tessellation, which is a partition of a plane (or a planar domain) into  
the union of convex  polygons $V_i$, called Voronoi cells, 
corresponding to the set of vertices ${\bf x}^i$. 
A Voronoi cell $V_i$ consists of all points in the plane 
which are closer to ${\bf x}^i$ 
than to any other vertex ${\bf x}^j, j \ne i$.

The edges of $V_i$ can lie either on $\partial \Omega$ or in
the interior of $\Omega$. On each face of $V_i$, that lies inside
$\Omega$,
$
\mid \ux - {\bf x}^i \mid = \mid \ux - {\bf x}^j \mid, \;\hbox{for
some}\; i \ne j.
$

\begin{definition}
\label{def:neighbors}
For each $i =
1,2 \ldots N$, define the index set ${\cal N}_i$ by
\[
{\cal N}_i = \left\{ j \in \{1, 2 \ldots, N\}, \; j \ne
i, \; \hbox{such that}\; V_i \; \hbox{and}\; V_j \; \hbox{have a
common edge} \right\}.
\]
We call a particle $D^j$ a {\bf neighbor} of $D^i$ if $j$ belongs to
${\cal N}_i$ that is the vertices ${\bf x}^i$ and ${\bf x}^j$ have a common edge in the 
Voronoi tessellation.  
\end{definition}
Note, that according to this definition, two particles are not neighbors if
their Voronoi cells have a common vertex but do not share an edge.

The minimal distance between neighboring particles $\DI$ and $D^j$
is given by
\begin{equation}
\delta_{ij} = \mid {\bf x}^i - {\bf x}^j \mid -2a.
\label{gap}
\end{equation}
We call $\delta_{ij}$ {\it inter-particle distances}.
If a disk $\DI$ is close to the external boundary $\bound$, we define
the particle-boundary minimal distance $\delta_i$ by
\begin{equation}
\label{bgap}
\delta^i={\rm dist}({\bf x}^i, \bound)-a.
\end{equation}

To model the high concentration regime, we assume that $\delta_{ij}$ and $\delta_{i}$ 
satisfy
\begin{equation}
\label{ip-dist}
\delta_{ij}=\delta d_{ij}, \;\;\;\;\;\delta_i=\delta d_i,
\end{equation}
where $\delta$ is a small parameter in the problem, and the 
{\it scaled inter-particle distances} $d_{ij}, d_i$ satisfy
\begin{equation}
\label{sc-dist}
c \leq d_{ij}\leq 1,\;\;\;\;c \leq d_{i}\leq 1,
\end{equation}
with a fixed positive $c$ independent of $i,j$. 

\begin{definition}
\label{def:network}
The centers ${\bf x}^i$ of the particles $D^i$ are called the 
{\bf interior vertices} of the network (graph) $\Gamma$, $i = 1, 2 \ldots N$.  
The {\bf interior edges}
$b_{ij}$ of the network connect neighboring vertices ${\bf x}^i$ and ${\bf x}^j$. If a
Voronoi cell $V_i$ has edges belonging to $\partial \Omega^+ \bigcup
\partial \Omega^-$, the corresponding ${\bf x}^i$ is called a {\bf near-boundary
vertex}. Each near-boundary vertex ${\bf x}^i$ is connected with
$\bound^\pm$
by an {\bf exterior edge}, which is a  segment $\tilde b_i$ perpendicular to $\bound^\pm$. 
This segment intersects $\bound^\pm$ at 
an {\bf exterior vertex} denoted by  $\tilde{\ux}^{(i)}$.

Finally, the {\bf network (graph)} $\Gamma$ is the collection of all interior vertices, exterior vertices, and all the edges connecting these vertices. 
\end{definition}

The set of indices of the near-boundary vertices will be denoted by $I$, that is
$i\in I$ if the vertex ${\bf x}^i$ is connected to $\bound$. Also, we write $i\in I^+(I^-)$
if ${\bf x}^i$ is connected to $\bound^+(\bound^-)$.

Note that $\Gamma$ is essentially the Delaunay graph dual to the Voronoi
tessellation, and the above notions admit straightforward generalization to 
three dimensions.

\section{Network approximation of the effective viscosity}
\subsection{Network equations}

To define the network approximation, we first assign
a translational velocity ${\bf T}^i$ and an angular velocity 
$\omega^i$ of a particle $D^i$ to the corresponding interior vertex ${\bf x}^i$.
At the exterior vertices $\tilde {\bf x}^i$ we prescribe
the velocity vector ${\bf g}$ which represents the boundary conditions
(\ref{shear-fbc}) or (\ref{ext-fbc}).

For each pair of neighboring particles $\DI$ and $D^j$ we introduce a gap $\Pi_{ij}$
which represents a fluid region where lubrication effects are very strong as shown on
Fig.1. The width $R_{ij}$ of such gap is independent of $\delta$. For technical
reasons it is convenient to work with non-intersecting gaps. Since the maximal number
of gaps adjacent to each particle is equal to the maximal coordination number (number of neighbors),
K, we can choose $R_{ij}=KMa$, where $M$ is sufficiently small and fixed.
The precise value of $M$ is not essential for our
purposes.

\begin{figure}
\label{gap1}
\begin{center}
\input{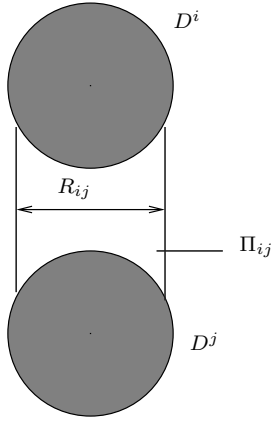}
\caption{A gap $\Pi_{ij}$ between the neighboring particles $D^i$ and $D^j$.}
\end{center}
\end{figure}

The orientation of each interior gap $\Pi_{ij}$ relative to
a disk $D^i$ is specified by a unit vector 
\begin{equation}
\label{orient}
{\bf q}^{ij}=\frac{{\bf x}^i-{\bf x}^j}{|{\bf x}^i-{\bf x}^j|}.
\end{equation}
We also let ${\bf p}^{ij}$ be the unit vector obtained by rotating
${\bf q}^{ij}$  clockwise by $\pi/2$ (see Fig.2).

Boundary edges and corresponding gaps are oriented perpendicular to ${\bf e}_1$. 
This 
reflects the physical fact that the zone of the largest energy
dissipation is located near the shortest line connecting ${\bf x}^i$ with
the boundary. Therefore, ${\bf q}^i={\bf e_2}$, (respectively $-{\bf e}_2$),
when $\DI$ is adjacent to $\bound^+$ (respectively $\bound^-$).

\begin{figure}
\label{gap2}
\begin{center}
\input{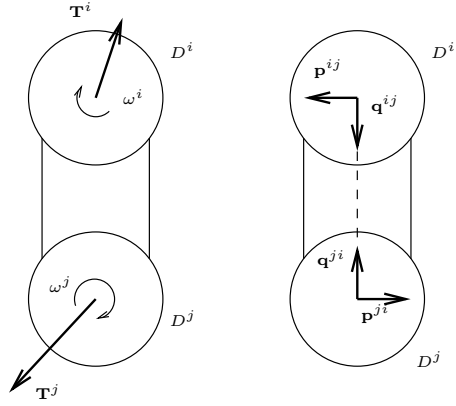}
\caption{Assignment of ${\bf T}^i, \omega^i$ and 
orientation of the gap between two neighboring particles.}
\end{center}
\end{figure}

Next, to each edge of the network we associate a dissipation rate
$W^{ij}$ ($W^i$), calculated in the corresponding gap $\Pi_{ij}$ ($\Pi_i$).
The calculation of the dissipation rates employs lubrication approximation in the gap.
The velocity in the gap is decomposed into three velocities, representing
the "elementary" motions 
called {\it spring motion}, shear,  and rotation
(see Fig. 3). The total velocity field in a gap is the sum of
these elementary velocities and a "residual" velocity field, whose contribution
to the gap dissipation rate is $O(1)$ as $\delta\to 0$.  Lubrication approximations
for each of the elementary velocities and estimates for the residual
are obtained in \cite{BBP} where more details can be found.

\begin{figure}
\label{gap2}
\begin{center}
\includegraphics{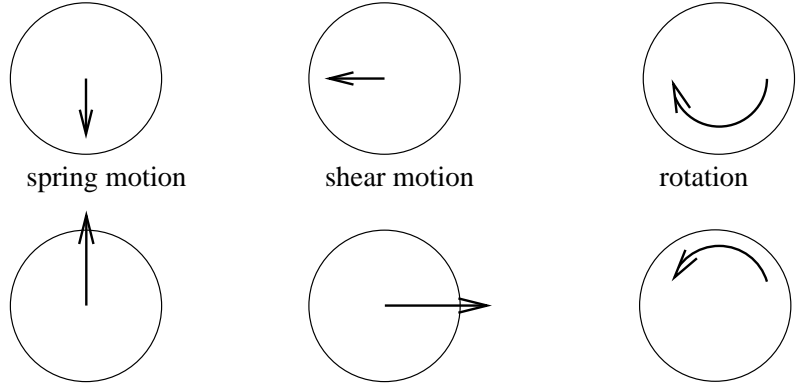}
\caption{Three elementary motions. Arrows represent the boundary conditions.}
\end{center}
\end{figure}

Using approximations of the elementary velocities to calculate (up to the terms of order
$O(1)$ as $\delta\to 0$) the
dissipation rates in each gap we obtain
\begin{equation}
\label{loc1-dis}
\begin{array}{ccc}
W^{ij} & = & \hspace{-5.1cm}\delta^{-3/2} C^{ij}_{sp}\left[(\TI-\TJ) \cdot {\bf
q}^{ij}\right]^2+ \\
& & \delta^{-1/2} C^{ij}_{sh}\left[(\TI -\TJ)\cdot
{\bf p}^{ij} + a{\om}^i + a{\om}^{j} \right]^2 +
\delta^{-1/2}C^{ij}_{rot}a^2(\omega^i-\omega^j)^2,\\
\end{array} 
\end{equation}
in the interior gaps $\Pi_{ij}$,
and 
\begin{equation}
\label{loc1-dis}
W^{i}=\delta^{-3/2} C^{i}_{sp}\left[(\TI-{\bf
g})\cdot{\bf q}^{i} \right]^2 +\delta^{-1/2} C^{i}_{rot}a^2\left(2 {\om}^i\right)^2+ 
\delta^{-1/2} C_{sh}^i \left[(\TI -{\bf g})\cdot {\bf p}^{i} + a{\om}^i
\right]^2 
\end{equation}
in the boundary gaps $\Pi_i$. 
The expressions for factors $C^{ij}_{sp}, C^{ij}_{sh}$ and
$C^{ij}_{rot}$  and $C^i_{sp}, C^i_{sh}, C^i_{rot}$ are calculated explicitly
in \cite{BBP}:   
\begin{equation}
\label{net-2D-4}
\begin{array}{ccccccc}
C^{ij}_{sp} &=& \frac 34 \pi\mu \big(\frac{a}{d_{ij}}\big)^{3/2}+
\frac{27}{10} \pi\mu \big(\frac{a}{d_{ij}}\big)^{1/2},& \hspace{0.5cm}&
C^{i}_{sp} &=& \frac 34 \pi\mu \big(\frac{a}{d_{i}}\big)^{3/2}+
\frac{27}{10} \pi\mu \big(\frac{a}{d_{i}}\big)^{1/2},
\\
C^{ij}_{sh} & =& \frac 12 \pi\mu \big(\frac{a}{d_{ij}}\big)^{1/2},\hspace{1.4cm}&
\hspace{0.5cm} &
C^{i}_{sh} & =& \frac 12 \pi\mu \big(\frac{a}{d_{i}}\big)^{1/2},\hspace{1.4cm}
\\
C^{ij}_{rot} & =& \frac{9}{16} \pi\mu \big(\frac{a}{d_{ij}}\big)^{1/2}.\hspace{1.4cm}&
\hspace{0.5cm}&
C^{i}_{rot} & =& \frac{9}{16} \pi\mu \big(\frac{a}{d_{i}}\big)^{1/2}.\hspace{1.4cm}
\\
\end{array}
\end{equation}
In the formulas (\ref{net-2D-4}), $d_{ij}, d_i$ are the scaled inter-particle distances 
defined in (\ref{sc-dist}). 

The sum of the local dissipation rates $W^{ij}, W^i$
is a quadratic form 
\begin{equation}
\begin{array}{lll}
Q & = & \sum_{\Pi_{ij}} W^{ij} +\sum_{\Pi_i} W^i\\
 &= &  {\displaystyle
\sum_{i=1}^N} {\displaystyle \sum_{{\tiny \begin{array}{c}j \in {\cal
N}_i\\ j < i
\end{array}}}}\left\{
\delta^{-3/2} C^{ij}_{sp}\left[(\TI-\TJ) \cdot {\bf
q}^{ij}\right]^2+\delta^{-1/2} C^{ij}_{sh}\left[(\TI -\TJ)\cdot
{\bf p}^{ij} + a{\om}^i + a{\om}^{j} \right]^2 \right.  \\ &&
+\left. \delta^{-1/2} C^{ij}_{rot}a^2\left({\om}^i -
{\om}^j\right)^2 \right\} + {\displaystyle
\sum_{i \in I}} \left\{ \delta^{-3/2} C^{i}_{sp}\left[(\TI-{\bf
g})\cdot{\bf q}^{i} \right]^2 +\delta^{-1/2} C^{i}_{rot}a^2\left(2 {\om}^i\right)^2 \right.  
\\ 
&& + \left. \delta^{-1/2} C_{sh}^i \left[(\TI -{\bf g})\cdot {\bf p}^{i} + 
a{\om}^i
\right]^2 \right\},
\end{array}
\label{funct-q}
\end{equation}
where $I$ denotes the set of indices of the near-boundary vertices defined in
Section 4.

The form $Q$ is the discrete network approximation of the functional $W_{\Omega_F}$ in 
the variational principle (\ref{primal}). 
The main idea of the network approximation is that the most of the energy is dissipated
in the gaps $\Pi_{ij}, \Pi_i$, so that 
\begin{equation}
\label{E}
E=E_{net}+O(1),\;\;\;\;{\rm as}\;\delta\to 0.
\end{equation}
The discrete dissipation rate $E_{net}$ in (\ref{E}) is 
defined by
\begin{equation}
\label{ddrate}
E_{net}=\min_{{\bf T}^i, \omega^i} Q=Q({\bf T}_{min}^1, {\bf T}^2_{min},...
{\bf T}^N_{min}, {\om}^1_{min},...{\om}^N_{min}),
\end{equation}
where the vectors $\TI_{min}, i=1,2,\ldots, N$ and scalars ${\om}^i, i=1, 2, \ldots, N$ minimize 
$Q$ in (\ref{funct-q}). The minimum in (\ref{ddrate}) is taken over all possible
collections of ${\bf T}^i, \omega^i$.

It is well known that solving the minimization problem for $Q$ is
equivalent to solving the linear system  (Euler-Lagrange equations), which is obtained
by equating the gradient of $Q$ to zero. Setting to zero partial derivatives with respect to
$T^i_l, l=1,2$ we obtain
\begin{eqnarray}
\sum_{j \in {\cal N}_i} \left\{ \delta^{-3/2} C^{ij}_{sp}
\left[(\TI-\TJ)\cdot{\uq}^{ij}\right] {\uq}^{ij}\right\}+\hspace{3.65cm} 
\nonumber\\
\sum_{j \in {\cal N}_i} \left\{\delta^{-1/2} C^{ij}_{sh}\left[(\TI-\TJ) \cdot{\up}^{ij}+ a{\om}^i +
a{\om}^j \right] {\up}^{ij} \right\} +{\bf B}^i = {\bf F}^i, ~ ~
\label{bal-force}
\end{eqnarray}
for each $i=1,2,..., N$,
where
\begin{eqnarray}
{\bf B}^i &=& \left\{ \begin{array}{l}
\delta^{-3/2} C^{i}_{sp}(\TI \cdot {\bf
q}^i) {\bf q}^i 
+ \delta^{-1/2} C^{i}_{sh} \left[
\TI\cdot {\bf p}^i + a {\om}^i\right] {\bf p}^i, ~
~\mbox{if} ~i \in I \\ 
{\bf 0} ~ ~\mbox{otherwise},
\end{array} \right. 
\label{bd-force} \\
\end{eqnarray}
\begin{eqnarray}
{\bf F}^i &=& \left\{ \begin{array}{l}
\delta^{-3/2} C^{i}_{sp}({\bf g} \cdot {\bf
q}^i) {\bf q}^i 
+ \delta^{-1/2} C^{i}_{sh} (
{\bf g}\cdot {\bf p}^i) {\bf p}^i, ~
~\mbox{if} ~i \in I \\ 
{\bf 0} ~ ~\mbox{otherwise},
\end{array} \right. 
\label{rhs-force} \\
\end{eqnarray}
Next, equating the partial derivatives $\frac{\partial Q}{\partial \omega^i}$ to zero we obtain
\begin{eqnarray}
\sum_{j\in
{\cal N}_i} \left\{ \delta^{-1/2} C^{ij}_{sh}\left[(\TI-\TJ)\cdot{\up}^{ij}+ a{\om}^i +
a{\om}^j\right]\right\} + \nonumber \\
\sum_{j\in
{\cal N}_i} \left\{ \delta^{-1/2} C^{ij}_{rot}\left({\om}^i -
{\om}^j\right)\right\} +  {\cal B}^i = {\cal M}^i,\hspace{1.5cm}
\label{bal-torque}
\end{eqnarray}
for all $i = 1, \ldots, N$, where
\begin{eqnarray}
{\cal B}^i &=& \left\{ \begin{array}{l}
\delta^{-1/2} C_{sh}^i \left(\TI \cdot {\bf
p}^i +a {\om}^i\right) + 4 \delta^{-1/2} C^{i}_{rot} {\om}^i, ~
~\mbox{if} ~i \in I, \\ 0 ~ ~\mbox{otherwise}. 
\end{array}\right.
\label{bd-torque} 
\end{eqnarray}
\begin{eqnarray}
{\cal M}^i &=& \left\{ \begin{array}{l}
\delta^{-1/2} C_{sh}^i \left({\bf g} \cdot {\bf
p}^i\right) ~
~\mbox{if} ~i \in I, \\ 0 ~ ~\mbox{otherwise}. 
\end{array}\right.\hspace{4cm}
\label{torque-rhs} 
\end{eqnarray}

Equations (\ref{bal-force}) and (\ref{bal-torque}) are, respectively,
the equations of force and torque balance of the particles,
and the minimization in (\ref{funct-q}) ensures that the rigid body
translational and angular velocities are chosen in such a way that
the suspension is in mechanical equilibrium. Note also that (\ref{bal-force})
is a system of $2N$ equations, and (\ref{bal-torque}) is a system of
$N$ equations. Together they form $3N$ equations for $3N$ unknowns
$({\bf T}^i, \omega^i)$. The coefficients and right hand side of
(\ref{bal-force}) are of order $\delta^{-3/2}$ and $\delta^{-1/2}$ while
all the coefficients in (\ref{bal-torque}) are of order $\delta^{-1/2}$.
When all ${\bf T}^i$ are zero (no translations), the remaining terms
are of order $\delta^{-1/2}$, but in the case $\omega^i=0, i=1,2,\ldots, N$
(no rotations), the remaining equations contain terms of order $\delta^{-3/2}$.
This reflects the well known fact that the contributions from local translational spring motions are stronger
than the contributions from rotational and other translational motions. Keeping only
spring translations we obtain
the truncated (leading order) discrete dissipation functional
\begin{equation}
\label{lead-funct}
\delta^{-3/2}\widehat Q=
\delta^{-3/2} \displaystyle{\left[\frac 12 \sum_{i=1}^N \sum_{j\in {\cal N}_i} C^{ij}_{sp}
(({\bf T}^i-{\bf T}^j)\cdot{\bf q}^{ij})^2+\sum_{i\in I} C_{sp}^{i}
((\TI-\ug)\cdot {\bf q}^{i})^2\right].}
\end{equation}
Then $Q$ can be decomposed as follows. 
\begin{equation}
\label{q-decomp}
Q= \delta^{-3/2}\widehat Q+ \delta^{-1/2} Q^\prime,
\end{equation}
where the coefficients of the forms $\widehat Q$ and $Q^\prime$ do not depend on 
$\delta$.

We next show that the total discrete dissipation rate $E_{net}$ can be estimated by the truncated
dissipation rate obtained by minimizing $\widehat Q$. 
Let $\widehat{\TI}, i=1,2,\ldots, N$
denote the translational velocities which minimize $\widehat Q$ ( they are clearly 
independent
of $\delta$). Since $Q^\prime$ in (\ref{q-decomp}) is non-negative, 
\begin{eqnarray}
\delta^{-3/2}\widehat Q(\widehat{\TI})\leq \delta^{-3/2}\widehat Q(\TI_{min}, 
{\om}^i_{min})\leq
E_{net}=\nonumber \hspace{1cm}\\
Q(\TI_{min}, \omega^i_{\min})\leq \delta^{-3/2}\widehat 
Q(\widehat{\TI})+\delta^{-1/2}Q^\prime(\widehat{\TI},{\om}^i=0).
\label{var-in}
\end{eqnarray}
Since $Q^\prime(\widehat{\TI},{\om}^i=0)\leq C$ with $C$ independent of 
$\delta$ as $\delta\to 0$,
(\ref{var-in}) implies 
\begin{equation}
\label{lead-term}
E_{net}=\delta^{-3/2}\widehat E+O(\delta^{-1/2}),\;\;\;\;{\rm as}~\delta\to 0,
\end{equation}
where
\begin{equation}
\label{lead-dis}
\widehat E=\widehat Q(\widehat{\TI})={\displaystyle \min_{{\tiny
{\bf T}^1,\ldots {\bf T}^N}} \widehat Q ({\bf T}^1,\ldots {\bf T}^N}).
\end{equation}
This equation enables one to calculate the leading term in the asymptotics
of the effective viscosity by solving a simplified minimization problem
involving only the translational particle velocities. However, this algorithm
is useful only when 
\begin{equation}
\label{crucial}
\widehat E>0,
\end{equation}
because in this case the leading term in the asymptotics of the dissipation rate is of order
$\delta^{-3/2}$.
If $\min\widehat Q=0$, the leading term degenerates, and 
the rate of blow up in (\ref{lead-dis}) is at most $\delta^{-1/2}$.

Minimization of the truncated quadratic form $\widehat Q$ corresponds to solving
the truncated linear system of the Euler-Lagrange equations
\begin{equation}
\label{lead-system}
\sum_{j\in {\cal N}_i}\big[C^{ij}_{sp} 
(\widehat{{\bf T}}^i-\widehat{{\bf T}}^j)\cdot{\uq}^{ij}{\uq}^{ij}\big]+
{\bf B}({\widehat{\TI}})={\bf R}^i,\;\;\;\;i=1,2, \ldots, N,
\end{equation}
where
\begin{equation}
\label{B}
{\bf B}(\widehat{\TI})=
\begin{cases}
C_{sp}^{i}
(\widehat{\TI}\cdot {\bf q}^{i}){\bf q}^{i},\;\;\;\;{\rm when}~i \in I&\\
0 \;\;\;\;\;\;{\rm otherwise}.&\\
\end{cases}
\end{equation}
The right hand side vectors ${\bf R}^i$ represent external boundary conditions:
\begin{equation}
\label{lead-rhs}
{\bf R}^i=
\begin{cases}
C_{sp}^{i}
({\bf g}\cdot {\bf q}^{i}){\bf q}^{i},\;\;\;\;{\rm when}~i \in I&\\
0 \;\;\;\;\;\;{\rm otherwise}.&\\
\end{cases}
\end{equation}

To better see the structure of the functional $\widehat Q$ and the system (\ref{lead-system}) 
it is convenient to rewrite them
in a compact form as 
\begin{equation}
\label{comp-hatq}
\hat Q({\bf z})=\frac 12 A{\bf z}\cdot{\bf z}-{\bf f}\cdot {\bf z}+r,
\end{equation}
and
\begin{equation}
\label{compls}
A{\bf z}={\bf f},
\end{equation}
where  ${\bf z}$ is the particle velocity vector, and ${\bf f}$ is the vector of discretized
boundary conditions. The vectors     
${\bf z}, {\bf f}\in {\bf R}^{2N}$ are defined by
\begin{equation}
\label{zf}
 {\bf z}=
\left(
\begin{array}{c}
 T^1_1\\
 T^1_2\\
 T^2_1\\
 T^2_2\\
  ...\\
 T^N_1\\
 T^N_2\\
\end{array}
\right), 
\;\;\;\;\;\;\;{\rm and}\;\;\;\;\;\;\;
{\bf f}=2
\left(
\begin{array}{c}
 R^1_1\\
 R^1_2\\
 R^2_1\\
 R^2_2\\
  ...\\
 R^N_1\\
 R^N_2,\\
\end{array}
\right).  
\end{equation}
The fixed scalar $r$ equals $\tilde {\bf f}\cdot {\bf f}$ where components of 
$\tilde{\bf f}$ are
defined by the components of ${\bf f}$, as follows.  
$$
\tilde f_{2k-1}=\frac{1}{4 C_{sp}^k}f_{2k-1},\;\;\;\;
\tilde f_{2k}=\frac{1}{4 C_{sp}^k}f_{2k}\;\;\;\;\;\;k=1,2,\ldots, N.
$$
The matrix $A$ in (\ref{comp-hatq}) is symmetric and non-negative definite. 
Its entries are determined by the scaled inter-particle distances $d_{ij}$,
particle radius $a$ and vectors 
${\bf q}^{ij}, {\bf q}^i$ defined by (\ref{orient}).
The vector ${\bf f}$ and the scalar $r$ in (\ref{comp-hatq}) are determined by the 
boundary conditions on $\bound^+,\bound^-$ (extensional or shear). The linear term
${\bf f}\cdot{\bf z}$ depends on the translational velocities of the particles located
near $\bound^+$ or $\bound^-$.

To the leading order in $\delta$, the effective viscosity is determined by the
discrete dissipation rate $\delta^{-3/2}\widehat E$ where $\widehat E$
is the minimum of 
$\widehat Q$. Thus the qualitative behaviour of the effective viscosity is 
determined
by solving (\ref{compls}). We now sketch the issues which arise in computing the leading
term of the effective viscosity. First, in the case of shear viscosity,
${\bf f}=0$, which means that the shear boundary conditions do not contribute
to the strong blow up. This results in the weak blow up of $\mu^\star$ (see section 6).
In case of extensional conditions, ${\bf f}\ne 0$, which means that the right hand side
of the full network equations (\ref{bal-force}) contains terms of order $\delta^{-3/2}$.
In this case, calculation of the leading term is impeded by the fact that the matrix
$A$ is not invertible. Indeed,
from (\ref{lead-funct}) it is clear that the value of $\widehat Q$ does not 
change when all vectors ${\bf T}^i, i=1,2, \ldots, N$ are replaced by
$\TI+t{\bf e}_1$ (horizontal translation by
$t{\bf e}_1$, $t$  arbitrary real). This is not surprising, 
because the functional $\widehat Q$ 
is invariant under horizontal translation of ${\bf T}^i$. The invariance is due
to the vertical orientation of the boundary gaps 
explained in section 5.1.
Note that this is not equivalent to translation of a coordinate system, since ${\bf g}$
is not changed. Translational invariance of $\widehat Q$ implies that 
any ${\bf z}^*$ that solves the non-homogeneous system (\ref{compls}) produces other solutions
of the form 
$
{\bf z}^*+t{\bf w}_0,
$
where $t$ is arbitrary real, and
\begin{equation}
\label{w-not}
{\bf w}_0=(1,0, 1, 0, \ldots, 1,0)^T.
\end{equation}
This means that vectors of the form $t{\bf w}_0$ solve the homogeneous system 
\begin{equation}
\label{comphom}
A{\bf z}={\bf 0}.
\end{equation}
Because of the above mentioned invariance of
the functional $\widehat Q$, the leading term in the asymptotics of effective
viscosity will be uniquely defined, unless (\ref{comphom}) has
some other nontrivial solutions. Therefore, it makes sense to look
for conditions on the network which would guarantee that every solution of the
homogeneous system is of the form $t{\bf w}_0$. Then the 
non-homogeneous system (\ref{compls}) would be uniquely solvable
up to horizontal translation.
In Section 7.3 we show that for a typical random distribution of particles
this is indeed the case.

The last issue concerns the validity of the estimate (\ref{crucial}). The functional
$\widehat Q$ is non-negative, but it may be zero.  When (\ref{crucial})
holds, local lubrication analysis provides the correct order of the leading
term in the asymptotics of the extensional effective viscosity
($\delta^{-3/2}$ in dimension two and $\delta^{-1}$ in dimension three). 
If (\ref{crucial})
does not hold, that is,
\begin{equation}
\label{anticrucial}
\min \widehat Q=0,
\end{equation}
then the leading term in dimension two is of order $\delta^{-1/2}$, ($\ln(1/\delta)$ in 
dimension three). 

Whether or not
the estimate (\ref{crucial}) holds, depends on the geometry of the 
particle array as well as
the boundary conditions on $\bound^+$ and $\bound^-$. 
In section 7.4 we show that in the case of extensional boundary conditions,
(\ref{crucial}) holds for generic arrays, and thus the leading
term in the asymptotic of $E_{net}$ ( and extensional effective viscosity
$\lambda^*$, see (\ref{eev1})) is of the order $\delta^{-3/2}$. 
However,  there exist special arrays for which the extensional effective viscosity
is of order $\delta^{-1/2}$. An example of such an array is presented in section 7.5.

\section{Effective shear viscosity}
In this Section we show that in dimension two, the asymptotic order of the shear 
effective viscosity $\mu^\star$ is $\delta^{-1/2}$, while
the local lubrication analysis predicts
the rate $\delta^{-3/2}$. In three
dimensions, $\delta^{-1/2}$ and $\delta^{-3/2}$ should be replaced by, 
respectively, 
$\ln\delta$ and $\delta^{-1}$ (see \cite{BBP}). 
The local analysis in three dimensions predicts that the asymptotics of the 
shear effective viscosity $\mu^\star$ (see (\ref{shev2}))  
should be of order $\delta^{-1}$, but
numerical simulations in \cite{Brady} and experimental results
in \cite{ShP} and \cite{VW} show that random suspensions in shear flow have effective viscosity
of order $\ln \delta$.  Our estimate $\mu^\star=O(\delta^{-1/2})$ is therefore
in agreement with the three-dimensional results in \cite{Brady},
up to the difference in the dimension of the space. 
 
The decrease in the asymptotic order of $\mu^\star$ is a global
phenomenon, which shows that
the local analysis could be misleading, and analysis of the entire
particle array leads to qualitatively different results.
This fact can be explained as follows.
The "strong" blow up rate ($\delta^{-1}$ in three dimensions and $\delta^{-3/2}$ in two
dimensions) is obtained using classical lubrication techniques applied to two
particles $D^i, D^j$ whose translational and angular velocities $\TI, \TJ, \omega^i, \omega^j$
are {\it independent of $\delta$}. However, 
the network analysis of the ensemble of particles, 
interacting with each other and with the external boundary, shows that the above velocities 
may depend on $\delta$. In the case of shear boundary conditions, the right hand side of the 
network equations (\ref{bal-force}), (\ref{bal-torque}) is 
of order $\delta^{-1/2}$, whereas the matrix of the network
equations is of order $\delta^{-3/2}$. The solution of the network equations
is thus small (of order $\delta$) which makes local and global dissipation rates small. 

When the boundary conditions are given by (\ref{shear-fbc}), the vectors ${\bf R}^i$ in
(\ref{lead-rhs}) are zero (since ${\bf q}^i\cdot{\bf e}_1=0$ for all $i=1,2, \ldots, N$). 
Consequently,
the right hand side ${\bf f}$ in (\ref{compls}) is zero. The functional $\widehat Q$ reduces to
$A{\bf z}\cdot {\bf z}$ which is clearly zero for every solution of the homogeneous system
$A{\bf z}=0$. Since ${\bf z}=0$ is an admissible trial vector for $\widehat Q$, 
$\widehat E=\min \widehat Q=0$, and from (\ref{lead-term}) we obtain
\begin{equation}
\label{shear-diss}
\mu^\star= C E_{net}=O(\delta^{-1/2}),
\end{equation}
where $C=2\gamma^{-2}|\Omega|^{-1}$ (see (\ref{shev2})).
The same conclusion can be obtained by directly estimating the minimum of
the form 
$Q$. Consider first the shear boundary conditions (\ref{shear-fbc}) with $\gamma=1$
and a simple example of a two-disk network on Fig. 4.
\begin{figure}
\label{chain}
\begin{center}
\includegraphics{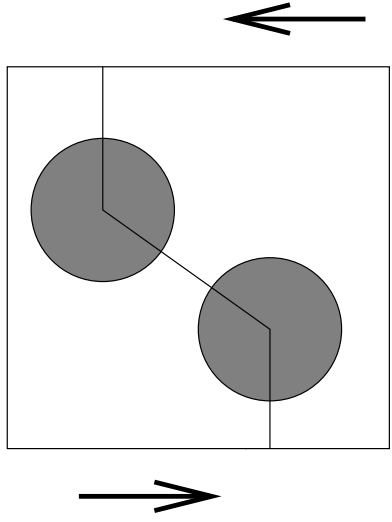}
\caption{A two-disk network with shear boundary conditions.}
\end{center}
\end{figure}
The functional $Q$ for this example has the form
\begin{equation}
\label{shearex1}
\begin{array}{ccc}
Q &= & \hspace{-0.3cm}\delta^{-3/2}\left\{
C_{sp}^{12}\left[({\bf T}^1-{\bf T}^2)\cdot {\bf q}^{12}\right]^2+
C_{sp}^1\left[({\bf T}^1-{\bf e}_1)\cdot {\bf e}_2\right]^2+
C_{sp}^2\left[({\bf T}^2+{\bf e}_1)\cdot {\bf e}_2\right]^2\right\}\\
&&\hspace{-0.65cm} +\delta^{-1/2}\left\{C_{sh}^{12}\left[({\bf T}^1-{\bf T}^2)\cdot {\bf p}^{12}+a\omega_1+
a\omega_2\right]^2+C_{rot}^{12}a^2(\omega_1-\omega_2)^2\right\}\hspace{2cm}\\
&&\hspace{-2.3cm} +\delta^{-1/2}\left\{C_{sh}^1\left[({\bf T}^1-{\bf e}_1)\cdot {\bf e}_1+a\omega_1\right]^2+
C_{sh}^2\left[({\bf T}^2+{\bf e}_1)\cdot {\bf e}_1+a\omega_2\right]^2\right\}\\ 
& &\hspace{-7.25cm}+\delta^{-1/2}\left\{4C_{rot}^1a^2{\omega_1}^2+
4C_{rot}^2a^2{\omega_2}^2\right\}.
\end{array}
\end{equation}
The dissipation rate $E$ is the minimum of $Q$. Hence, for any collection
${\bf T}^1, {\bf T}^2, \omega_1, \omega_2$, we have $E\leq Q({\bf T}^1, 
{\bf T}^2, \omega_1, \omega_2)$.
In particular, choosing ${\bf T}^1={\bf T}^2={\bf 0}, \omega_1=\omega_2=0$ in 
(\ref{shearex1}) we obtain
$$
E\leq \delta^{-1/2}(C^1_{sh}+C^2_{sh})
$$
Since $C_{sh}^1, C_{sh}^2$ are independent of $\delta$, the blow up rate of $E$ is 
at most $\delta^{-1/2}$.

Next, consider the general case.
Since for shear boundary condition $\ug \bot {\bf q}^i$ for all $i\in I$,  we obtain
\begin{equation}
\label{gen-ex}
\begin{array}{lll}
Q &= &  {\displaystyle
\sum_{i=1}^N} {\displaystyle \sum_{{\tiny \begin{array}{c}j \in {\cal
N}_i\\ j < i
\end{array}}}}\left\{
\delta^{-3/2} C^{ij}_{sp}\left[(\TI-\TJ) \cdot {\bf
q}^{ij}\right]^2\right\}+\\
& & 
{\displaystyle
\sum_{i=1}^N} {\displaystyle \sum_{{\tiny \begin{array}{c}j \in {\cal
N}_i\\ j < i
\end{array}}}}\left\{\delta^{-1/2} C^{ij}_{sh}\left[(\TI -\TJ)\cdot
{\bf p}^{ij} + a{\om}^i + a{\om}^{j} \right]^2 \right. 
+\left. \delta^{-1/2} C^{ij}_{rot}a^2\left({\om}^i -
{\om}^j\right)^2 \right\} + \\
& & {\displaystyle
\sum_{i \in I}} \left\{ \delta^{-3/2} C^{i}_{sp}\left[\TI\cdot{\bf q}^{i} 
\right]^2 +\delta^{-1/2} C^{i}_{rot}a^2\left(2 {\om}^i\right)^2 \right\} + \\
& & 
{\displaystyle
\sum_{i \in I}}\left\{\delta^{-1/2} 
C_{sh}^i \left[(\TI\cdot {\bf p}^{i}-{\bf g}\cdot{\bf p}^i + 
a{\om}^i
\right]^2 \right\},
\end{array}
\end{equation}

Choosing the trial vectors $\TI$ as $\TI=0, i=1,2, \ldots, N$ and $\omega^i=0$, $i=1,2, \dots, N$
we obtain
\begin{equation}
\label{sp-en-shear}
E_{net}=\min Q\leq Q({\bf T}^i={\bf 0}, \omega^i=0)=\delta^{-1/2} \sum_{i\in I}
C_{sh}^{i}(\ug\cdot {\bf p}^{i})^2 =\gamma^2 \delta^{-1/2}\sum_{i \in I}C_{sh}^i.
\end{equation}
where we used the shear boundary conditions (\ref{shear-fbc}).
The sum in the right hand side of (\ref{sp-en-shear}) 
is independent of $\delta$ and the shear rate $\gamma$. 
This estimate shows that under the boundary conditions 
(\ref{shear-fbc}),
the effective viscosity blows up as $\delta^{-1/2}$ at most.
More precisely, we have the following\newline
\noindent
{\bf Conclusion.} {\it Effective shear
viscosity of a concentrated suspension 
has the blow up rate $\delta^{-1/2}$ in two dimensions, that is 
$$
\mu^\star\leq C \delta^{-1/2}\;\;\;{\rm as}~\delta\to 0,
$$
with $C$ independent of $\delta, \gamma$.}

\section{Extensional effective viscosity}
\subsection{Simplification of boundary conditions and outline of the method}
In this Section we show that for the extensional boundary conditions, the leading term
in the asymptotics of the dissipation rate $E$ from (\ref{lead-term}) may or may not be zero 
depending on the geometry of
a particle array, that is, the leading term in the asymptotics of the effective
extensional viscosity is either of order $\delta^{-3/2}$ (strong blow up) or $\delta^{-1/2}$
(weak blow up). We provide two geometric conditions on the network graph which ensure
strong blow up. 

In a planar steady extensional flow of the effective fluid, the rate of strain
tensor is
\begin{equation}
\label{eff-ext-strain}
e({\bf v}^0)=
\left(
\begin{array}{cc}
\epsilon & 0\\
0 & -\epsilon\\
\end{array}
\right),
\end{equation}
where $\epsilon$ denotes a constant extension rate. The corresponding
velocity field is of the form
\begin{equation}
\label{ext-eff-vel}
{\bf v}^0=(\epsilon x_1, -\epsilon x_2)^T,
\end{equation}
which gives the boundary conditions
\begin{equation}
\label{ext-eff-bc}
{\bf v}^0= 
\begin{cases}
(\epsilon x_1, -\epsilon)^T, \;\;{\rm when}\;x_2=1 \;\;({\rm on}\;\partial\Omega^+),&\\
(\epsilon x_1, \epsilon)^T, \;\;{\rm when}\;x_2=-1 \;\;({\rm on}\;\partial\Omega^-).&\\
\end{cases}
\end{equation} 
We decompose ${\bf v}^0$ as 
\begin{equation}
\label{ext-vsplit}
{\bf v}^0={\bf v}^0_{vc}+{\bf v}^0_{ge},
\end{equation}
where
${\bf v}^0_{vc}$ is a vertical contraction velocity
satisfying
\begin{equation}
\label{contr-bc}
{\bf v}^0_{vc}={\bf g}_{vc}=
\begin{cases}
-\epsilon{\bf e}_2~~~{\rm on}~\bound^+, &\\
\;\;\;\epsilon {\bf e}_2~~~{\rm on}~\bound^-, &\\
\end{cases}
\end{equation}
and ${\bf v}^0_{ge}$ is the horizontal extension velocity field 
with the boundary conditions given by
\begin{equation}
\label{ext-bc}
{\bf v}^0_{ge}={\bf g}_{ge}
\begin{cases}
\epsilon x_1{\bf e}_1~~~{\rm on}~\bound^+, &\\
\epsilon x_1 {\bf e}_1~~{\rm on}~\bound^-. &\\
\end{cases}
\end{equation}
Since
\begin{equation}
\label{ext-gsplit}
{\bf g}={\bf g}_{vc}+{\bf g}_{ge},
\end{equation}
and ${\bf g}_{ge}\bot {\bf q}^i, i\in I$, the value of $\widehat Q$ in 
(\ref{lead-funct}) does not change when ${\bf g}$ in (\ref{lead-funct}) is replaced by
${\bf g}_{vc}$. Hence,
\begin{equation}
\label{q-vext}
Q({\bf T}^i, \omega^i, {\bf g})=\delta^{-3/2}\widehat Q({\bf T}^i,{\bf g}_{vc})+
\delta^{-1/2}Q^\prime({\bf T}^i, \omega^i, {\bf g}_{vc}+{\bf g}_{ge}),
\end{equation}
To determine the rate of blow up of the dissipation rate, we need
to analyze the minimizers of $\widehat Q({\bf T}^i,{\bf g}_{vc})$. The estimate (\ref{lead-term})
implies that the second term in the right hand side of (\ref{q-vext}) is of order $\delta^{-1/2}$ 
(at most). Since $\widehat Q$ is independent of $\delta$, its minimizing vectors
$\widehat \TI$ are also $\delta$-independent. Consequently, the blow up rate of the 
dissipation
depends on whether the minimum of $\widehat Q(\TI, {\bf g}_{vc})$ is positive. If it is, 
the extensional effective 
viscosity  $\lambda^\star$ is of order $\delta^{-3/2}$, otherwise $\lambda^\star$ grows no 
faster
than $\delta^{-1/2}$. Which type of behavior occurs, depends on the validity of the estimate
(\ref{crucial}). As mentioned in section 5, (\ref{crucial}) may fail for certain particle
arrays. In this section we provide geometric conditions  which insure positivity of 
$\min \widehat Q$, 
and give examples of networks for which this minimum is zero. The principal conclusion
here is that extensional viscosity of suspensions with a comparable volume fraction of particles
may vary by an order of magnitude in the inter-particle distance, depending
on the geometry of a particle array.

Our method of analysis is based on the following simple observation. The form $\widehat Q(\TI, {\bf g}_{vc})$
is a sum of non-negative terms, namely
\begin{equation}
\label{whq-simpl}
\widehat Q(\TI, {\bf g}_{vc})= 
\displaystyle{\frac 12 \sum_{i=1}^N \sum_{j\in {\cal N}_i} C^{ij}_{sp}
(({\bf T}^i-{\bf T}^j)\cdot{\bf q}^{ij})^2+\sum_{i\in I} C_{sp}^{i}
((\TI-{\bf g}_{vc})\cdot {\bf q}^{i})^2.}
\end{equation}
This shows that $\min \widehat Q(\TI, {\bf g}_{vc})=0$ if and only if the minimizing vectors
${\TI}, i=1,2, \dots, N$ satisfy the system of equations
\begin{equation}
\label{ns}
\begin{array}{c}
({\bf T}^i-{\bf T}^j)\cdot{\bf q}^{ij}=0,\;\;\;\;\;i=1,2..N, j\in {\cal N}_i,\\
\hspace{-2.2cm}({\TI}-{\bf g}_{vc})\cdot {\bf q}^{i}=0, \;\;\;\;\;i\in I.
\end{array}
\end{equation}
Hence, if (\ref{ns}) does not have solutions, (\ref{crucial}) must hold. 
Also, it should be noted that  if the estimate (\ref{crucial}) holds for
some subgraph of the network, then it also holds for the whole network.
This observation can be used to reduce the network to
a simpler graph, for which it is easier to prove (\ref{crucial}).

\subsection{Simple examples}
Suppose that the boundary conditions are given by (\ref{contr-bc}) with $\epsilon=-1$.
In this section we  present three simple examples of networks with small number of vertices. 
One of these networks
has $\widehat E=0$, and for the other two examples $\widehat E>0$. 

\noindent
{\bf Example 1.} This an example of the network for which $\widehat E=0$. 
To demonstrate this, we show that 
there is a nontrivial particle velocity vector ${\bf z}$ such that $\widehat Q({\bf z})=0$.
Consider the network on Fig. 5. 
\begin{figure}
\label{ex1}
\begin{center}
\includegraphics{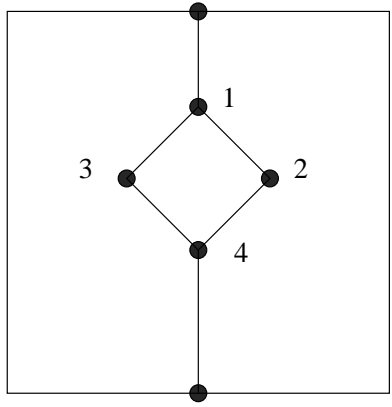}
\caption{A network of four vertices with $\widehat E=0$.}
\end{center}
\end{figure}
The vectors ${\bf q}^{ij}$ are defined as follows. 
\begin{equation}
\label{qi}
\begin{array}{cc}
\hspace{-3cm}{\bf q}^1={\bf e}_2, & \hspace{-4cm}{\bf q}^4=-{\bf e}_2, \\ 
{\bf q}^{12}={\bf q}^{34}=\frac{1}{\sqrt{2}}({\bf e}_1-{\bf e}_2),& \\
{\bf q}^{13}={\bf q}^{24}=-\frac{1}{\sqrt{2}}({\bf e}_1+{\bf e}_2) & \\
\end{array}
\end{equation}
Next, define ${\bf T}^i$ as follows.
\begin{equation}
\label{ti}
{\bf T}^1={\bf e}_2,\;\;{\bf T}^2=-{\bf e}_1,\;\;\;
{\bf T}^3={\bf e}_1,\;\;\;
{\bf T}^4=-{\bf e}_2.
\end{equation}
The functional $\widehat Q$ corresponding to this network is
\begin{equation}
\label{ex1q}
\begin{array}{ccc}
\widehat Q & = &C^1_{sp}\left[({\bf T}^1-{\bf e}_2)\cdot{\bf q}^1\right]^2+
C^{12}_{sp}\left[({\bf T}^1-{\bf T}^2)\cdot{\bf q}^{12}\right]^2+\\
 & & C^{13}_{sp}\left[({\bf T}^1-{\bf T}^3)\cdot{\bf q}^{13}\right]^2+
C^{24}_{sp}\left[({\bf T}^2-{\bf T}^{4})\cdot{\bf q}^{24}\right]^2+\\
 & & C^{34}_{sp}\left[({\bf T}^3-{\bf T}^4)\cdot{\bf q}^{34}\right]^2+
C^4_{sp}\left[({\bf T}^4+{\bf e}_2)\cdot{\bf q}^4\right]^2.
\end{array}
\end{equation}
When ${\bf T}^i$ are defined by (\ref{ti}), all the scalar product in brackets in (\ref{ex1q})
are zero, and therefore $\min\widehat Q=0$.\\
\noindent
{\bf Example 2.}  For the network of three vertices in Fig. 6, $\min \widehat Q>0$. To show this,
\begin{figure}
\label{ex1}
\begin{center}
\includegraphics{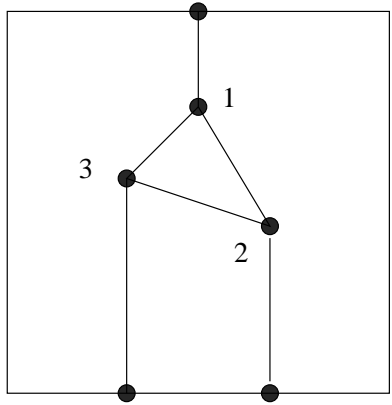}
\caption{A network of three vertices with $\widehat E>0$.}
\end{center}
\end{figure}
consider the system corresponding to the general system (\ref{ns}). 
\begin{equation}
\label{ex2.1}
\begin{array}{c}
({\bf T}^1+{\bf e}_2)\cdot {\bf e}_2=0,\;\;\;({\bf T}^1-{\bf T}^2)\cdot {\bf q}^{12}=0,\;\;\;\;
({\bf T}^1-{\bf T}^3)\cdot {\bf q}^{13}=0,\\
({\bf T}^2-{\bf T}^3)\cdot {\bf q}^{23}=0,
\;\;\;\;({\bf T}^2-{\bf e}_2)\cdot {\bf e}_2=0,\;\;\;\;\;
({\bf T}^3-{\bf e}_2)\cdot {\bf e}_2=0.\\
\end{array}
\end{equation}
We prove that the system (\ref{ex2.1}) has no solutions. Indeed, the last two equations
imply ${\bf T}^2=t_2{\bf e}_1+{\bf e}_2$, ${\bf T}^3=t_3{\bf e}_1+{\bf e}_2$, for some scalars
$t_2, t_3$. Next, the first equation in the second row of (\ref{ex2.1}) yields
$(t_2-t_3){\bf e}_1\cdot{\bf q}^{23}=0$, and thus $t_2=t_3=t$. Substituting 
${\bf T}^2={\bf T}^3=t{\bf e}_1+{\bf e}_2$ into the second
and third equations in the first row of (\ref{ex2.1}) we obtain
\begin{equation}
\label{ex2.2}
\begin{array}{c}
({\bf T}^1-t{\bf e}_1-{\bf e}_2)\cdot{\bf q}^{12}=0,\\
({\bf T}^1-t{\bf e}_1-{\bf e}_2)\cdot{\bf q}^{13}=0.\\
\end{array}
\end{equation}
Since ${\bf q}^{12}, {\bf q}^{13}$ are non-collinear, (\ref{ex2.2}) yields 
${\bf T}^1=t{\bf e}_1+{\bf e}_2$, which contradicts the first equation in the first row of 
(\ref{ex2.1}).\\
\noindent
{\bf Example 3.} Next, consider a rectangular network of four vertices in Fig. 7.
\begin{figure}
\label{ex1}
\begin{center}
\includegraphics{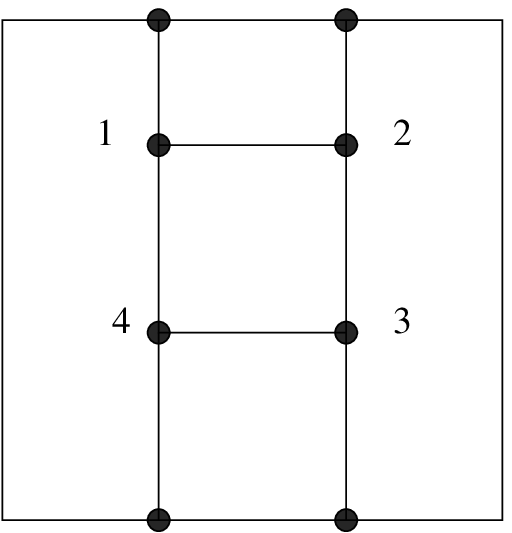}
\caption{A network of four vertices with $\widehat E>0$.}
\end{center}
\end{figure}
The system (\ref{ns}) for this example becomes
\begin{equation}
\label{ex3.1}
\begin{array}{c}
({\bf T}^1-{\bf T}^2)\cdot {\bf e}_1=0,\;\;\;\;
({\bf T}^1-{\bf T}^4)\cdot {\bf e}_2=0,\;\;\;\;
({\bf T}^2-{\bf T}^3)\cdot {\bf e}_2=0,\;\;\;\;
({\bf T}^3-{\bf T}^4)\cdot {\bf e}_1=0,\;\;\;\;\\
\hspace{-0.6cm}({\bf T}^1+{\bf e}_2)\cdot {\bf e}_2=0,\;\;\;\;\;\;
({\bf T}^2+{\bf e}_2)\cdot {\bf e}_2=0,\;\;\;\;
({\bf T}^3-{\bf e}_2)\cdot {\bf e}_2=0,\;\;\;\;
({\bf T}^4-{\bf e}_2)\cdot {\bf e}_2=0.\\
\end{array}
\end{equation}
The last two equations in the second row of (\ref{ex3.1}) yield ${\bf T}^3=t_3{\bf e}_1+{\bf e}_2$,
${\bf T}^4=t_4{\bf e}_1+{\bf e}_2$, with some scalars $t_3, t_4$. Next, the second and third 
equations in the first row of 
(\ref{ex3.1}) produce ${\bf T}^1=t_1{\bf e}_1+{\bf e}_2$, ${\bf T}^2=t_2{\bf e}_1+{\bf e}_2$,
which contradict, respectively, the first and second equations in the second row.

The three examples above seem to indicate 
that two basic building blocks for networks with $\widehat E>0$ (strong blow up) are
triangles (Fig. 6) and  rectangles aligned with the edges of $\Omega$ (Fig. 7). 
Misaligned rectangular structures such as shown in Fig. 8 would produce $\widehat E=0$ (weak blow up).
\subsection{Quasi-triangulated graphs}
\subsubsection{Definition of quasi-triangulated graphs. Solvability of the system (\ref{compls})}
The network $\Gamma$ partitions $\Omega$ into a a 
disjoint union of convex polygons, which are called
{\it Delaunay cells}. When points ${\bf x}^j$ are distributed randomly in  $\Omega$, the interior
Delaunay cells are 
typically triangles. This simple but important fact can be explained as follows.
The edges of Voronoi tessellation are 
perpendicular bisectors of the edges of Delaunay cells.
If an interior Delaunay cell is, for instance, a quadrilateral, then
any two vertices lying on a diagonal cannot be neighbors, and therefore 
the point of intersection of four edges of the Voronoi tessellation must
be equidistant from the four vertices. This means that a convex quadrilateral may be a Delaunay
cell only if all four vertices lie on a circle. When the vertices of the network
are randomly placed, the likelihood of four (or more) points lying on the same 
circle is small. It is natural to call such cells the defect cells.
Therefore, most of the interior Delaunay cells of a random
network are triangles. Other polygonal cells (quadrilateral, pentagonal etc.)
are typically small in number, isolated and are likely to be unstable in the actual flow.

Since the exterior edges of the network are vertical, the cells adjacent to the boundary
are typically quadrilateral. An example of a generic network  is shown on Fig. 8.

\begin{figure}
\label{del1}
\begin{center}
\includegraphics{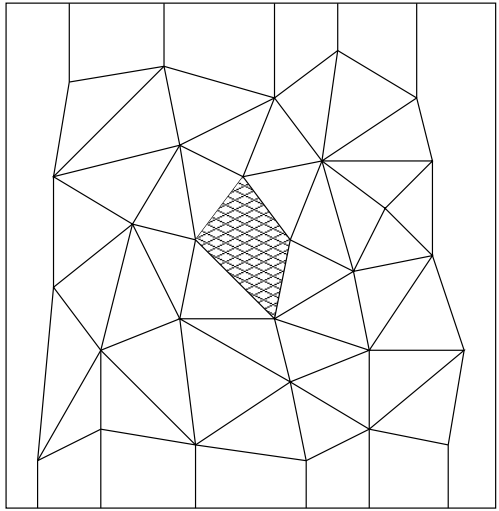}
\caption{An example of the network. The 
shaded quadrilateral represents a defect cell.}
\end{center}
\end{figure}

Next, we define a broad class of graphs containing
generic Delaunay-type networks. The graphs in this class are
called {\it quasi-triangulated}, because of the presence of triangular cells. 
The number of triangular cells may be relatively small, as indicated by the examples
below. We show that for quasi-triangulated graphs the leading term in the asymptotics
can be effectively computed by minimizing the functional $\widehat Q$, that is the
minimum is unique and can be found by solving the linear system (\ref{compls}).
We also show that the crucial estimate (\ref{crucial}) holds even for more general
graphs which contain a "spanning" quasi-triangulated subgraph.

Given an arbitrary network graph $\Gamma$, we define its 
maximal quasi-triangulated 
subgraph $\Gamma_M$ by the following 
iterative procedure.\newline
\noindent
{\it Step 1.} Consider interior vertices which are connected to 
$\bound^-$ and call these vertices {\it generation one} vertices. All interior edges 
connecting these vertices are {\it generation one} edges. Add all generation one
edges and vertices to the subgraph.\newline
\noindent
{\it Step 2.} Consider all remaining vertices which  are connected to the vertices
of the subgraph
by at least two non-collinear edges. These vertices and edges belong to generation 
two. Add them to the subgraph. Note that the non-collinearity condition
leads to formation of "supportive triangles". \newline
\noindent
{\it Step 3.} Repeat step 2 until no more vertices can be added.

If the maximal quasi-triangulated subgraph $\Gamma_M$
contains all interior vertices of $\Gamma$, we call the graph $\Gamma$ 
{\it quasi-triangulated}. It turns out that for quasi-triangulated graphs
the system (\ref{compls}) is "almost uniquely" solvable, that is
two solutions differ by a vector of the form $t{\bf w}_0$, where ${\bf w}_0$ is defined by (\ref{w-not}), and
the value of the form $Q$ for these solutions is the same.

\begin{proposition}
\label{quasi}
Suppose that the network graph $\Gamma$ is quasi-triangulated. Then there is a
unique solution of the system (\ref{compls}), up to a horizontal translation.
\end{proposition}

This proposition is proved in Appendix B.

\subsubsection{Examples of quasi-triangulated graphs}
First we observe that a restriction to $\Omega$ of a periodic rectangular lattice  is not 
quasi-triangulated. By contrast, a periodic triangular lattice restricted to $\Omega$ is 
quasi-triangulated.
(see Fig. 9).

\begin{figure}
\label{del1}
\begin{center}
\includegraphics{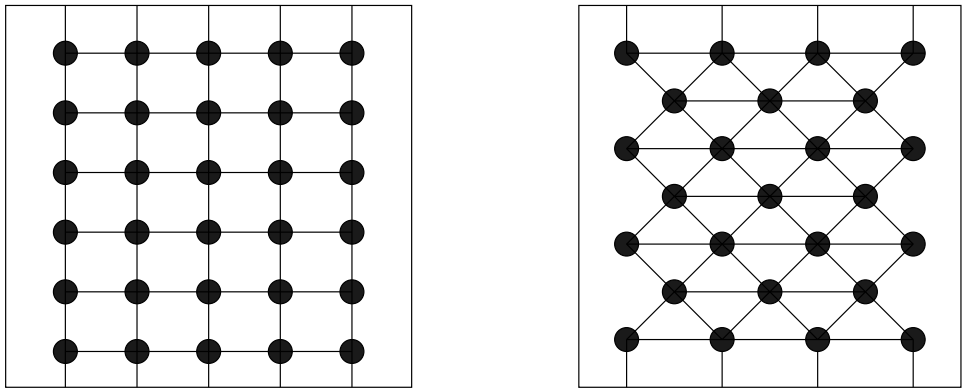}
\caption{A rectangular graph on the left is not quasi-triangulated. Clearly, in this case
$\Gamma_F=\Gamma^-$.
The triangular graph on the right
is quasi-triangulated.}
\end{center}
\end{figure}

In fact, if a network $\Gamma$ is not periodic, but all of its interior Delaunay cells 
are triangles, 
then $\Gamma$ is quasi-triangulated.
The converse is false (see an example in Fig. 11). This means that the 
quasi-triangulation property is 
more general than triangulation property. In \cite{BBP} we considered a class of graphs
containing a triangulated path (see Fig. 10). In Fig 11, we present two examples 
showing that a quasi-triangulated graph need not contain a triangulated path. On 
the other hand,
a graph containing a triangulated path is not necessarily quasi-triangulated.
However, when a triangulated path exists, it must be contained in the maximal
quasi-triangulated subgraph. Therefore, in this case the maximal
quasi-triangulated subgraph is spanning ("extends from top to bottom"). Below in section
7.4.1 we show that this condition is sufficient for positivity of the minimum
of $\widehat Q$. 

\begin{figure}
\label{del3}
\begin{center}
\includegraphics{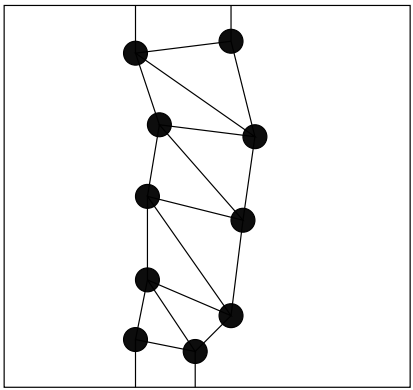}
\caption{A triangulated path.}
\end{center}
\end{figure}

\begin{figure}
\label{del2}
\begin{center}
\includegraphics{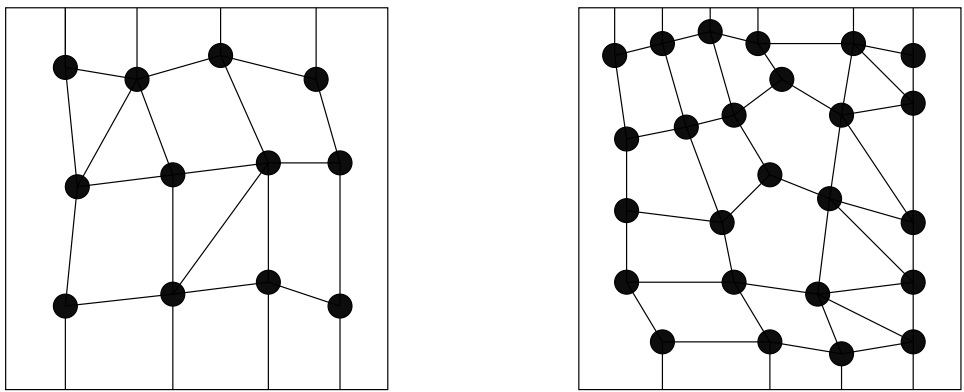}
\caption{The graph on the left is quasi-triangulated but does not contain a 
triangulated path. The graph
on the right contains a triangulated path, but is not quasi-triangulated.}
\end{center}
\end{figure}

\subsection{Strong blow up of $\lambda^\star$. Percolating rigidity networks}
\subsubsection{Quasi-triangulated subgraphs}

For the rest of this section, we  consider the steady flow of the suspension 
corresponding to 
the boundary conditions (\ref{contr-bc}) with the extension
rate $\ep=1$.

We shall say that a network  is a {\it percolating rigidity graph} when
the crucial estimate (\ref{crucial}) holds.
In this subsection we show that quasi-triangulated graphs are percolating rigidity graphs.
Moreover, a graph has percolating rigidity even when it is not
quasi-triangulated but contains a spanning quasi-triangulated subgraph.

\begin{proposition}
\label{tricon}
Suppose that the boundary conditions are given by (\ref{contr-bc}) and the network 
graph $\Gamma$ contains a spanning quasi-triangulated subgraph. Then $\Gamma$ is a percolating rigidity graph. Consequently, the
extensional effective viscosity  $\lambda^\star$ of suspensions corresponding to such networks
is $O(\delta^{-3/2})$. 
\end{proposition}

The proof of this proposition is given in Appendix C.

\subsubsection{Networks containing a vertical path. Periodic square networks}
We now present another class of percolating rigidity graphs. Namely, these
are graphs that contain a path connecting $\bound^+$ and $\bound^-$, such that
all edges in this path are oriented along ${\bf e}_2$-direction
(vertical). The simplest representative of this class of graphs is a periodic square
lattice. 
A periodic square graph does not contain a spanning quasi-triangulated
subgraph (in this case $\Gamma_M$ is just a path
which consists of the vertices adjacent to $\bound^-$, see Fig. 9). However, the
estimate (\ref{crucial}) holds for a periodic square graph such as the one shown on Fig. 9.
This follows from  the more general criterion

\begin{proposition}
\label{cubic}
Suppose that a network graph $\Gamma$ contains a path $\Gamma^\pm$ such that

\noindent
i) it connects $\bound^+$ and $\bound^-$,

\noindent
ii) all edges of $\Gamma^\pm$ are vertical.

Then $\Gamma$ is a percolating rigidity graph.
\end{proposition}

As noted above, if a path $\Gamma^\pm$ has percolating rigidity, then
the "larger" graph $\Gamma$ is also a percolating rigidity graph. Therefore, we only need
to show that a path of vertical edges has percolating rigidity, that is, the quadratic form
$\widehat Q$ corresponding to such path is positive-definite. 

For simplicity, consider a 
path containing three 
vertices. The argument can be directly generalized to an arbitrary number of vertices.
As before, (\ref{crucial}) will hold if the corresponding system (\ref{ns}) has 
no solutions. The latter now has the form
\begin{equation}
\label{vpath1}
\begin{array}{c}
({\bf T}^1-{\bf e}_2)\cdot{\bf e}_2=0,\\
({\bf T}^1-{\bf T}^2)\cdot{\bf e}_2=0,\\
({\bf T}^2-{\bf T}^3)\cdot{\bf e}_2=0,\\
({\bf T}^3+{\bf e}_2)\cdot{\bf e}_2=0,\\
\end{array}
\end{equation}
Introduce new unknown vectors $\widehat {\bf T}^i={\bf t}^i+{\bf e}_2$ $i=1,2,3$.
The last equation of (\ref{vpath1}) yields $\widehat {\bf T}^3=t_3 {\bf e}_1$,
where $t_3$ is a scalar. Then from the third equation we obtain $\widehat {\bf T}^2=
t_2{\bf e}_1$, and then the second equation yields $\widehat{\bf T}^1=t_1{\bf e}_1$, which
contradicts the first equation. The contradiction shows that the system (\ref{vpath1})
has no solutions. The argument can be easily modified to show that if at least one
of the edges of a path is non-vertical, then this path does not have percolating rigidity.

Proposition \ref{cubic} shows that for rectangular lattices oriented
parallel to the edges of $\bound$, asymptotics of extensional effective viscosity 
is of order $\delta^{-3/2}$ which corresponds to strong blow up.
Asymptotic formulas in \cite{Keller} also predict strong blow up
of the extensional viscosity for cubic lattices. Thus our results are consistent
with the results in \cite{Keller}. We refer to section 8 for a more detailed
comparison.

\subsection{Weak blow up}
In this section we present an example of a particle array for which
the leading term in the asymptotics of $\lambda^\star$ is zero.
Roughly speaking, the array in question is a rectangular lattice
rotated so that none of its interior edges are vertical. 
Let ${\bf k}$ denote a unit vector non-collinear to either ${\bf e}_1$ or ${\bf e}_2$. The interior
edges of the rectangular network in Fig. 12 are either parallel or perpendicular
to ${\bf k}$, while the prescribed boundary velocities are parallel to 
${\bf e}_2$.
This misalignment will lead to weak blow up.
To show that
the system (\ref{ns}) is solvable, we 
first consider a single path connecting $\bound^+$ and $\bound^-$.
This path may be any of the three such paths in Fig. 12. 
The argument we use admits a straightforward generalization to
a network with an arbitrary number of vertices.
\begin{figure}
\label{chain}
\begin{center}
\includegraphics{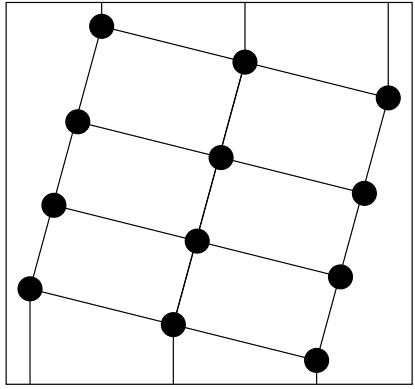}
\caption{A rotated rectangular lattice of 12 vertices.}
\end{center}
\end{figure}

The system (\ref{ns}) written for the path has the form
\begin{equation}
\label{npath1}
\begin{array}{c}
({\bf T}^1-{\bf e}_2)\cdot{\bf e}_2=0,\\
({\bf T}^1-{\bf T}^2)\cdot{\bf k}=0,\\
({\bf T}^2-{\bf T}^3)\cdot{\bf k}=0,\\
({\bf T}^3-{\bf T}^4)\cdot{\bf k}=0,\\
({\bf T}^4+{\bf e}_2)\cdot{\bf e}_2=0,\\
\end{array}
\end{equation}
For technical reasons it is convenient to introduce new unknowns ${\bf u}_1={\bf T}^1-{\bf e}_2$, 
${\bf u}_{12}={\bf T}^1-{\bf T}^2$, ${\bf u}_{23}={\bf T}^2-{\bf T}^3$,
${\bf u}_{34}={\bf T}^3-{\bf T}^4$. The relations between ${\bf u}_{ij}$ and
${\bf T}^i$ are 
\begin{eqnarray}
\label{relate}
{\bf T}^1={\bf u}_1+2{\bf e}_2,\nonumber\\
{\bf T}^2={\bf u}_1+2{\bf e}_2-{\bf u}_{12},\nonumber\\
{\bf T}^3={\bf u}_1+2{\bf e}_2-{\bf u}_{12}-{\bf u}_{23},\nonumber\\
{\bf T}^4={\bf u}_1+2{\bf e}_2-{\bf u}_{12}-{\bf u}_{23}-{\bf u}_{34}\label{last}.
\end{eqnarray}
From the first four equations of (\ref{npath1}) we see that ${\bf u}_1=t_1{\bf e}_1$, and 
${\bf u}_{i, i+1}=t_{i,i+1}{\bf k}^\bot$, $i=1,2,3$, where
${\bf k}^\bot$ denotes a unit vector orthogonal to ${\bf k}$, and $t_1, t_{i,i+1}$ are
scalars. The problems of solving
(\ref{npath1}) is now reduced to finding $t_1, t_{i,i+1}$ such that the fifth equation
of (\ref{npath1}) is satisfied. The fifth equation shows that 
\begin{equation}
\label{t4}
{\bf T}^4=t_4{\bf e}_1-{\bf e}_2,
\end{equation} 
where $t_4$ is a scalar. Equating (\ref{t4}) and (\ref{last}) we obtain the equation
for $t_1, t_{12}, t_{23}, t_{34}$ and $t_4$:
\begin{equation}
\label{ts}
t_4{\bf e}_1=t_1{\bf e}_1+{\bf e}_2-(t_{12}+t_{23}+t_{34}){\bf k}^\bot.
\end{equation}
This yields two scalar equations
\begin{equation}
\label{first}
0=1-(t_{12}+t_{23}+t_{34}){\bf k}^\bot\cdot{\bf e}_2,
\end{equation}
and
\begin{equation}
\label{second}
t_1-t_4=(t_{12}+t_{23}+t_{34}){\bf k}^\bot\cdot{\bf e}_1.
\end{equation}
The system of two equations (\ref{first}),(\ref{second}) for five unknowns has infinitely many
nontrivial solutions as long as ${\bf k}^\bot\cdot {\bf e}_2\ne 0$ , that is, the interior 
edges of the path are non-vertical.

At the next step of construction, we consider the full lattice on Fig. 6 containing 12 vertices.
The interior edges of the graph are oriented either by the unit vector ${\bf k}$ as above
(longitudinal edges), or
by ${\bf k}^\bot$ (latitudinal edges). 
We view the graph as the union of three paths of longitudinal edges 
extending from $\bound^-$ to $\bound^+$, 
with latitudinal edges connecting
these paths. To obtain the desired example, we choose the vectors ${\bf T}^i$ for one
of the paths, as explained above. Then we prescribe the same vectors to the corresponding
vertices of two remaining paths. Now, if two neighbors ${\bf x}^i$ , ${\bf x}^j$ belong
to different paths, then ${\bf T}^i={\bf T}^j$, so the corresponding equation
$({\bf T}^i-{\bf T}^j)\cdot{\bf k}^\bot=0$ of the system (\ref{ns}) is satisfied. When the neighbors
${\bf x}^i, {\bf x}^j$ belong to the same path, the corresponding equation 
$({\bf T}^i-{\bf T}^j)\cdot{\bf k}=0$ is satisfied by the choice of ${\bf T}^i, {\bf T}^j$.
Therefore, the whole system (\ref{ns}) in this case has infinitely many nontrivial solutions.
Each of these solutions makes $\widehat Q$, and thus the leading term in the asymptotics of
$\lambda^\star$, zero.

\section{Comparison with some results for periodic cubic  arrays in dimension three}
The main objective of network approximation is to define effective
properties for non-periodic deterministic or random arrays. While techniques of
periodic homogenization are well developed (\cite{BLP}, \cite{SP}, \cite{JKO} and references therein), non-periodic geometries are much less understood.
In the end of this section, we compare our results applied in particular case
of a periodic square array (in dimension two) with the results of \cite{Keller} 
obtained for cubic arrays in dimension three.

The effective viscosity of an infinite periodic suspension obtained in 
\cite{Keller}
is the fourth order tensor ${\mbox{\boldmath $\mu$}}^\star$ (in this section we use the
notation from from \cite{Keller}). 
In an effective flow
with the constant strain rate $\gamma$, the effective stress is
\begin{equation}
\label{k1}
S_{ij}^0=2\mu^\star_{ijkl}\gamma_{kl}-P\delta_{ij},
\end{equation} 
where $P$ is an effective pressure. The authors of \cite{Keller} obtained the following formula
for the components of 
${\mbox{\boldmath $\mu$}}^\star$:  
\begin{equation}
\label{visc-tensor}
\mu^\star_{ijkl}=\frac 12 \mu (1+\beta)(\delta_{ik}\delta_{jl}+\delta_{il}\delta_{jk}-
\frac{2}{3}\delta_{ij}\delta_{kl})+\mu(\alpha-\beta)(\delta_{ijkl}-
\frac 12 \delta_{ij}\delta_{kl}),
\end{equation}
where $\delta_{ijkl}=1$ if all indices are equal, otherwise $\delta_{ijkl}=0$.
$\mu$ is the fluid viscosity, and $\alpha, \beta$ are functions of the small
parameter $\epsilon$, related to the inter-particle distance $\delta$ as follows:
\begin{equation}
\label{del-ep}
\epsilon=\frac{\delta}{2a+\delta}\approx \frac{\delta}{2a},\;\;\;{\rm as}\;\delta\to 0.
\end{equation}
When the effective flow is incompressible, $\gamma_{ii}=0$, so the formula (\ref{visc-tensor})
simplifies to
\begin{equation}
\label{vtensor}
\mu^\star_{ijkl}=\frac 12 \mu (1+\beta)(\delta_{ik}\delta_{jl}+\delta_{il}\delta_{jk})+
\mu(\alpha-\beta)\delta_{ijkl}.
\end{equation}

For simple cubic lattices, 
up to the terms of order $O(1)$ in $\epsilon$, (\cite{Keller})
\begin{equation}
\label{alpha}
\alpha=\frac{3}{16}\pi \epsilon^{-1}+\frac{27}{80}\pi\ln \epsilon^{-1},
\end{equation}
and
\begin{equation}
\label{beta}
\beta=\frac 14 \pi\ln \epsilon^{-1}.
\end{equation}

Suppose that the imposed effective flow is a steady shear with the velocity
${\bf v}=(\kappa x_3, 0, 0)$ (three-dimensional analogue of (\ref{shear-v})), 
where $\kappa>0$ is a constant shear rate.
The components of the corresponding strain rate tensor  ${\mbox{\boldmath $\gamma$}}$ are
$\gamma_{13}=\gamma_{31}=\kappa>0$ and $\gamma_{ij}=0$ for other values of $(i,j)$.
Then, using (\ref{vtensor}), we obtain from (\ref{k1}) that the only nonzero
components of the effective deviatoric stress 
$2{\mbox{\boldmath $\mu$}}^\star{\mbox{\boldmath $\gamma$}}$ are
\begin{equation}
\label{k2}
(2\mu^\star\gamma)_{13}=(2\mu^\star\gamma)_{31}=2\mu (1+\beta)\kappa.
\end{equation}
Since $\alpha$ is not present in (\ref{k2}), the nonzero components of the deviatoric
effective stress are of order $\ln\epsilon^{-1}$ and thus the shear effective viscosity
calculated by the three-dimensional analogue of 
definition (\ref{shev2}) is of order $\ln\epsilon^{-1}$ (weak blow up in dimension three). The weak blow up was not identified in \cite{Keller}, but
it can be easily deduced from the formulas derived there.

In the case of an extensional flow, velocity vector
${\bf v}=(\kappa x_1, \kappa x_2, -2\kappa x_3)$, (compare with (\ref{ef-v-ext})),
where $\kappa>0$ is a constant extension
rate. The strain rate tensor is
\begin{equation}
\label{ext-st-3d}
{\mbox{\boldmath $\gamma$}}=
\left(
\begin{array}{ccc}
\kappa & 0 & 0\\
0 & \kappa & 0\\
0 & 0 & -2\kappa \\
\end{array}
\right)
\end{equation}
Then, using (\ref{k1}) and (\ref{vtensor}) we obtain the deviatoric stress
\begin{equation}
\label{ext-stress}
2{\mbox{\boldmath $\mu$}}^\star{\mbox{\boldmath $\gamma$}}=
\left(
\begin{array}{ccc}
L  & 0 & 0\\
0 & L & 0\\
0 & 0 & -2L, \\
\end{array}
\right)
\end{equation}
where 
\begin{equation}
\label{L}
L=2\kappa\mu(1+\beta)+\kappa\mu(\alpha-\beta).
\end{equation}
Components of the deviatoric stress contain $\alpha$ and are therefore of order $\ep^{-1}$. Consequently, the 
extensional effective viscosity is of order $\ep^{-1}$
(strong blow up in dimension three).

Although Noonan and Keller did not address the issue of weak versus strong blow up
for the effective viscosity in \cite{Keller}, the formulas (\ref{alpha})- (\ref{L}) 
are consistent with the results for square lattices in Section 6
of this paper in the following sense. If a periodicity cell corresponding to a simple 
cubic lattice is subjected to a uniform shear (extensional) flow, 
then the straightforward calculation presented above shows that the 
shear (extensional) effective
viscosity exhibits weak (strong) blow up. However, for other lattice types such
as BCC and FCC, formulas from \cite{Keller} imply strong blow up of both
viscosities, while our approach leads to the weak blow up of the shear viscosity
for all two-dimensional lattices. Perhaps, this can be attributed to the fact
that in \cite{Keller} effective viscosity is defined in a different way, as
explained in the introduction. In particular, our definition takes into account
boundary effects which are known to be essential in rheological measurements
(\cite{VW}, \cite{ShP}).
Considerations based on infinite periodic lattices cannot capture these boundary
effects. Futhermore, if our approach is applied to a rectangular periodic lattice
in a rectangular domain chosen so that lattice periodicity is preserved, then
our results are consistent with \cite{Keller}.
\section{Conclusions}
We have studied properties of the network approximation
of the shear effective viscosity
$\mu^\star$ and extensional effective viscosity $\lambda^\star$ of two-dimensional
flows of concentrated suspensions
with complex geometry. The flow domain was chosen to be a square, upper
and lower sides of which represented the physical flow boundary, while
the other two sides were assumed to be free surfaces. 

A small inter-particle
distance parameter $\delta$ was used to describe the high concentration regime
for particle arrays which are not necessarily periodic (i. g. random). 
In the recent paper \cite{Brady} by Sierou and Brady,
the high frequency dynamic viscosity of concentrated 
suspensions
(see \cite{VW}) was calculated as a function of volume fraction $\phi$ by means of accelerated Stokesian dynamics
simulations. Numerical results indicate
a singular behaviour of the effective viscosity as $\phi$ approaches
the maximal close packing fraction $\phi_{rcp}$.  We quote here from (\cite{Brady}):
''... The exact form of this singular behaviour
is not known. Results from lubrication theory for cubic lattices would suggest
that the singular form should consist both of $1/\epsilon$ and $\ln\epsilon$, where
$\ep=1-(\phi/\phi_{rcp})^{1/3}$, but the relative amount of each term is unknown...
As far as we are able to tell at this point, the $\ln\epsilon$ behaviour accurately describes
the numerical data.'' 

When the volume fraction is (approximately) constant over the flow region, $\ep\approx \delta$.
One of the objectives of our investigation
was to address the issue of the unexpectedly weak blow up in 
\cite{Brady}, and determine the asymptotic order of the effective
viscosity coefficients as $\delta\to 0$.
Our analysis of the shear viscosity $\mu^\star$, based on the discrete network approximation, 
showed that $\mu^\star=O(\delta^{-1/2})$ as $\delta\to 0$, while
the formal estimate based of the local lubrication analysis would give 
a higher rate $O(\delta^{-3/2})$.   
In dimension three, the corresponding rates, given by the network approximation, are, respectively, $\ln\delta$ and $\delta^{-1}$ (see \cite{BBP}).
Our analysis offers an explanation of the weak blow up of the shear viscosity. This analysis is also consistent with
the numerical simulations in \cite{Brady} up to the difference in
the space dimension.

We also show that the asymptotic order of the extensional viscosity $\lambda^\star$ depends on the geometry
of the particle array. For generic disordered arrays the network partitions the domain into 
polygons (Delaunay cells), most of which are triangles.  We have showed that for these generic 
arrays $\lambda^\star=O(\delta^{-3/2})$. The same asymptotic rate is
obtained for a larger class of networks called quasi-triangulated.
In a quasi-triangulated network, the percentage of triangular cells may be relatively small, but
the subnetwork containing triangular cells must be spanning. Another class of networks for which
$\lambda^\star=O(\delta^{-3/2})$ consists of rectangular periodic arrays aligned with the boundary of the flow. More
generally, the same rate is obtained for arrays containing a single spanning chain
of neighboring particles, perpendicular to the part of the boundary where the velocity is prescribed.

We show that in case of the strong blow up,
the leading term in the 
asymptotics of $\lambda^\star$ can be uniquely determined by solving a simplified linear system of the 
network equations, provided the array is quasi-triangulated. The simplified system, obtained by neglecting rotational and pairwise shear motions of particles,
provides an efficient computational tool for evaluating the dependence of the
effective viscosity on the geometry of the particle array and external boundary
conditions. 
When a network contains neither a spanning triangulated subgraph, nor a spanning vertical
path, the extensional effective viscosity can be of order 
$\delta^{-1/2}$. This weak blow up
is obtained, for instance, in the case of a periodic rectangular lattice, with interior edges misaligned with the orientation of the prescribed boundary data.

Our results imply that  the ratio of $\lambda^\star$
to $\mu^\star$ for generic disordered particle arrays is $O(\delta^{-1})$. By contrast,
in a Newtonian fluid this ratio depends only on the space dimension.
Therefore, our results indicate possible non-Newtonian behaviour of the effective fluid. 
\section{Acknowledgments}  
The authors wish to thank Professor John F. Brady for bringing the work \cite{Brady} to their attention. Work of Leonid Berlyand was supported in part by NSF grant DMS-0204637. Work of Alexander Panchenko was supported in part by the ONR grant N00014-001-0853.


\appendix
\section{Shear and extensional flows. Ratio of the viscosities in a 
Newtonian fluid}
\label{Appendix  A}
The following types of flows are relevant to our investigation.

\noindent
{\it Shear flow.} Consider the steady shear flow a homogeneous fluid
characterized by the constant shear rate $\gamma$. The velocity is given by
\begin{equation}
\label{shear-v}
{\uv}^0_{sh}=
\left(
\begin{array}{c}
\gamma x_2\\
0
\end{array}
\right).
\end{equation}
and the strain rate tensor is
\begin{equation}
\label{sh-strain}
{\ee}_{sh}^0=\frac{1}{2}
\left(
\begin{array}{cc}
0 & \gamma \\
\gamma & 0
\end{array}
\right).
\end{equation} 
Since the stress tensor is symmetric and independent of $\ux$, 
\begin{equation}
\label{eo-shear}
E^0=\int_\Omega \ts^0 \cdot {\ee}_{\it sh}^0d{\ux}=\frac 12 S^0_{12}\gamma |\Omega|,
\end{equation}
where $|\Omega|=\int_{\Omega} d\ux$. 

In the case of a homogeneous Newtonian fluid with viscosity $\mu$, $\ts^0=2\mu {\ee}_{sh}^0-PI$,
so that $S^0_{12}=\mu\gamma$. Using the formula (\ref{shev1}) we obtain
\begin{equation}
\label{newtmu}
\mu^\star=\mu,
\end{equation}
as expected.

\noindent
{\it Extensional flow.} In this case the fluid is being extended
in the horizontal direction and simultaneously contracted in the vertical direction at 
the same constant rate $\vep$. The velocity is
\begin{equation}
\label{ef-v-ext}
{\uv}^0_{\it ext}=
\left(
\begin{array}{c}
\vep x_1\\
-\vep x_2
\end{array}
\right),
\end{equation}
and
\begin{equation}
\label{ext-strain}
{\ee}_{ext}^0=
\left(
\begin{array}{cc}
\vep & 0 \\
0 & -\vep
\end{array}
\right).
\end{equation} 
For a homogeneous Newtonian fluid, $\ts^0_{ext}=2\mu {\ee}^0_{ext}-PI$, and thus
$S^0_{11}=2\mu\vep-P$, $S^0_{22}=-2\mu\vep-P$. Next using the definition (\ref{ev})
we obtain
\begin{equation}
\label{newtev}
\lambda^\star=4\mu.
\end{equation}
Therefore, for a Newtonian effective fluid the ratio $\lambda^\star/\mu^\star$ equals $4$ 
(in two dimensions). 
\section{Proof of Proposition \ref{quasi}}
\label{Appendix B}
The proposition \ref{quasi} will be proved if we prove the following

\begin{proposition}
\label{homsolve}
Suppose the network graph $\Gamma$ is quasi-triangulated. Then every solution of the 
homogeneous
system (\ref{comphom}) is of the form $t{\bf w}_0$ where $t$ is arbitrary 
real and ${\bf w}_0$
is given by (\ref{w-not}).
\end{proposition}
\noindent
{\it Proof.} 
First note that (\ref{comphom}) is Euler-Lagrange system for the functional
\begin{equation}
\label{q-hom}
Q_{hom}=\frac 12 A{\bf z}\cdot{\bf z}=
\displaystyle{\frac 12 \sum_{i=1}^N \sum_{j\in {\cal N}_i} C^{ij}_{sp}
(({\bf T}^i-{\bf T}^j)\cdot{\bf q}^{ij})^2+\sum_{i\in I} C_1^{i}
(\TI\cdot {\bf q}^{i})^2.}
\end{equation}
Clearly the minimum of $Q_{hom}$ is zero. Thus every solution of (\ref{comphom}) is a 
minimizer
of $Q_{hom}$. On the other hand, 
$Q_{hom}({\bf T}^1,\ldots {\bf T}^N)=0$ if and only if
the vectors $\TI$ satisfy the system of equations
\begin{equation}
\label{imp}
\begin{array}{cc}
({\bf T}^i-{\bf T}^j)\cdot{\bf q}^{ij}=0,&~~~~i=1,2\ldots N,~~j\in {\cal N}_i\\
\hspace{-11mm}\TI\cdot {\bf q}^{i}=0, & \hspace{-21mm}i\in I.
\end{array}
\end{equation}
Therefore, a vector ${\bf z}=({\bf T}^1, \ldots {\bf T}^N)^T$ solves (\ref{comphom})
if and only if $\TI, i=1,\ldots, N$ solve (\ref{imp}). 
The solvability of (\ref{imp}) will be directly linked to
the geometric structure of the graph $\Gamma$. 
We begin by observing that ${\bf q}^i=\pm{\bf e}_2$. Thus the second set of equations
in (\ref{imp}) yields 
\begin{equation}
\label{e1}
\TI=t^i {\bf e}_1, i\in I,
\end{equation}
that is, $\TI$ are horizontal
for all boundary vertices. Next, consider boundary vertices ${\bf x}^i,  i\in I^-$ 
(these are vertices connected to $\bound^-$), and recall that they belong to a path
$\Gamma^-$, edges of which are interior edges of $\Gamma$. 
Hence, if $i_1\in I^-$, then there is at least one 
$i_2\in I^-, i_2\ne i_1$ such that ${\ux}^{i_1}$ and ${\ux}^{i_2}$
are connected by an {\it interior} edge $b^{i_1i_2}$. 
Using the first set of equations in (\ref{imp}) and (\ref{e1}) we obtain
\begin{equation}
\label{e2}
({\bf T}^{i_1}-{\bf T}^{i_2})\cdot 
{\bf q}^{i_1i_2}=(t^{i_1}-t^{i_2}){\bf e}_ 1\cdot{\bf q}^{i_1i_2}=0.
\end{equation}    
Furthermore, $b^{i_1i_2}$ is non-vertical, that is, 
${\bf q}^{i_1i_2}\cdot {\bf e}_1\ne 0$, which yields
$t^{i_1}=t^{i_2}$. Since each ${\bf x}^i, i\in I^-$ is 
connected to at least one other, we obtain
\begin{equation}
\label{i-minus}
\TI=t{\bf e}_1, ~~~~i\in I^-,
\end{equation}
with the same scalar $t$. 

Since the graph is quasi-triangulated, there 
exists an {\it interior}
vertex  ${\ux}^{l_1}\in \Gamma$ , ${\ux}^{l_1}\notin \Gamma^-$, connected 
to at least two
vertices ${\ux}^{i_1}, {\ux}^{i_2},
i_1, i_2 \in I^-$ by non-collinear edges. Then, 
using the first set of equations in (\ref{imp})
we obtain
\begin{equation}
\label{e3}
\begin{array}{c}
({\bf T}^{l_1}-t{\bf e}_1)\cdot{\bf q}^{l_1,i_1}=0,\\
({\bf T}^{l_1}-t{\bf e}_1)\cdot{\bf q}^{l_1,i_2}=0.
\end{array}
\end{equation}
Since ${\bf q}^{l_1,i_1}$ and ${\bf q}^{l_1,i_2}$ are linearly independent, 
(\ref{e3}) yields
${\bf T}^{l_1}=t{\bf e}_1$. Next, let $G_1$ be the 
union of $\Gamma^-$,
${\ux}^{l_1}$ and all the edges which connect ${\ux}^{l_1}$ to $\Gamma^-$.
Using the definition of the quasi-triangulated graph again, we find a vertex ${\ux}^{l_2}$, not 
contained in 
$G_1$, and connected to $G_1$ by two non-collinear edges.
Repeating the argument following (\ref{e3}), we see that ${\bf T}^{l_2}=t{\bf e}_1$.
Then we choose $G_2$ to be the union of $G_1$, ${\ux}^{l_2}$ and all 
edges of $\Gamma$ which connect 
them. Repeating the process we find the 
vertex ${\ux}_{l_3}$ and continue until we obtain
\begin{equation}
\label{e4}
\TI =t{\bf e}_1, ~~~~~~~i=1, \dots N
\end{equation}
with the same scalar $t$. This means that every solution of (\ref{imp}) is of the form
$t{\bf w}_0$. Since solution spaces of (\ref{imp}) and (\ref{comphom}) are the same,
the proposition is proved.

The following discrete Korn inequality follows immediately from the Proposition 
\ref{homsolve}.

\begin{corollary}
Suppose that $\Gamma$ is quasi-triangulated.
Let $W\subset {\bf R}^{2N}$ be the one-dimensional subspace spanned by ${\bf w}_0$ from
(\ref{w-not}), and let $W^\bot$ denote the orthogonal complement of $W$ in ${\bf R}^{2N}$.
Also, let $Q_{hom}$ and $A$ be, respectively the quadratic form defined in (\ref{q-hom})
and its matrix. 
Then there is a constant $C>0$ such that the Korn-type inequality
\begin{equation}
\label{d-korn}
\frac 12 A{\bf z}\cdot{\bf z}=Q_{hom}({\bf z})\geq C {\bf z}\cdot{\bf z}
\end{equation}
holds for all ${\bf z}\in W^\bot$.
\end{corollary} 
Another straightforward corollary is as follows.
\begin{corollary}
\label{u-solve}
Suppose that $\Gamma$ is quasi-triangulated. Then the system
$$
A{\bf z}={\bf f}
$$
has a unique solution ${\bf z}\in W^\bot$ provided ${\bf f}\bot W$.
\end{corollary}

\noindent
{\bf Remark}. The projection $P_W$ onto the subspace $W$ is defined by
$$
P_W {\bf z}=\frac{{\bf z}\cdot{\bf w}_0}{N}{\bf w}_0.
$$
In terms of vectors $\TI$,
$$
P_W( {\bf T}^1,{\bf T}^2,\ldots {\bf T}^{N})=\frac{\sum_{i=1}^n T^i_1}{N}{\bf w}_0
$$
Therefore, using the definition of ${\bf f}$ in terms of ${\bf R}^i$, we can write the 
condition ${\bf f}\bot W$ as
\begin{equation}
\label{orthog}
\sum_{i=1}^n R^i_1=\sum_{i=1}^N{\bf R}^i\cdot {\bf e}_1=0
\end{equation}
 
The vectors ${\bf R}^i$ in (\ref{lead-rhs}) satisfy (\ref{orthog}), so that ${\bf f}$ in
(\ref{zf}) with $R^i$ defined by (\ref{lead-rhs}) is orthogonal to $W$. 
This gives the unique solvability of the network
equations (\ref{compls}).

\begin{corollary}
\label{al-solve}
Suppose that $\Gamma$ is quasi-triangulated.
Then there is a unique ${\bf z}^*\in W^\bot$ such that every solution of (\ref{compls})
is of the form ${\bf z}^*+t{\bf w}_0$, where $t{\bf w}_0\in W$.
\end{corollary} 
This means that solution of the network equations (\ref{compls}) is unique up to a 
horizontal translation.

\section{Proof of proposition \ref{tricon}}
\label{Appendix C}
\noindent
{\it Proof.}
First we observe that the form $\widehat Q$ is a sum of non-negative terms, each of 
which corresponds to an edge of the network graph $\Gamma$. Removal of an edge from 
$\Gamma$ corresponds to deletion of one non-negative term in $\widehat Q$. 
This means that for each subgraph $\Gamma^\prime$ of $\Gamma$, 
$\widehat Q(\Gamma)\geq \widehat Q(\Gamma^\prime)$. Next we choose $\Gamma^\prime$ to be
the maximal quasi-triangulated subgraph of $\Gamma$. 
We show that $\min \widehat Q(\Gamma^\prime)>0$. Indeed, $\min \widehat Q(\Gamma^\prime)=0$ 
if and only if
the corresponding system (\ref{ns}) has a solution.
To show that this system
has no solutions, introduce
new unknowns $\hat{\bf T}^i={\bf T}^i-{\bf e}_2$. Then from
(\ref{ns}) we obtain
\begin{equation}
\label{ns1}
(\hat{\bf T}^i-\hat{\bf T}^j)\cdot {\bf q}^{ij}=0,
\end{equation}
\begin{equation}
\label{ns2}
\hat{\bf T}^i\cdot {\bf q}^i=
\begin{cases}
-2\;\;\;\;{\rm when}\;i\in I^+&\\
0\;\;\;\;{\rm when}\;i\in I^-.&\\
\end{cases}
\end{equation}
Since the vectors ${\bf q}^i$ are vertical, (\ref{ns2}) yields
\begin{equation}
\label{bd} 
\hat{\bf T}^i=t^i {\bf e}_1,\;\;\;\;i \in I^-,
\end{equation}
where $t^i$ is a constant. 
Recall that $\Gamma$ contains a path $\Gamma^-$ which consists of all
boundary vertices connected to $\bound^-$ and all interior edges
connecting these vertices. Hence,
each ${\bf x}^{i_1}, i_1\in I^-$ has a neighbor ${\bf x}^{i_2},
i_2\in I^-$ and thus 
$(t^{i_1}-t^{i_2}){\bf e}_1\cdot{\bf q}^{i_1i_2}=0$ from (\ref{ns1}).
Since two boundary vertices cannot be joined by a vertical edge,
${\bf e}_1\cdot{\bf q}^{i_1i_2}\ne 0$. This implies that all
$t^i, i\in I^-$ are equal, that is 
\begin{equation}
\label{t}
\hat{\bf T}^i=t{\bf e}_1, i\in
I^-,
\end{equation}
where $t$ is a constant. 
Next, consider the boundary path $\Gamma^-$. By definition of $\Gamma^\prime$
there  is a vertex ${\bf x}^{j_1}, j_1 \not\in I^-$ connected
to the boundary vertices ${\bf x}^{i_1}, {\bf x}^{i_2}$, 
$ i_1, i_2 \in I^-$ by non-collinear edges of $\Gamma$.
Then from (\ref{ns1}) and (\ref{t}) we have
$$
\begin{array}{c}
(\hat{\bf T}^{j_1}-t{\bf e}_1)\cdot {\bf q}^{j_1,i_1}=0,\\
(\hat{\bf T}^{j_1}-t{\bf e}_1)\cdot {\bf q}^{j_1,i_2}=0.\\
\end{array}
$$
Since ${\bf q}^{j_1,i_1}$ and ${\bf q}^{j_1,i_2}$ are linearly
independent, we obtain $\hat{\bf T}^{j_1}=t{\bf e}_1$. Now this
argument can be used recursively. Next we choose $G_1$ to
be the union of vertices ${\bf x}^i, i\in I^-, {\bf x}^{j_1}$ and
the edges of $\Gamma^\prime$ which connect these vertices. Repeating the
argument, we find a vertex ${\bf x}^{j_2}\not\in G_1$,
connected to at least two vertices of $G_1$ by non-collinear
edges, which yields $\hat{\bf T}^{j_2}=t{\bf e}_1$, and so on, until
we obtain $\hat{\bf T}^i=t{\bf e}_1$ for all vertices ${\bf x}^i$ which belong to
$\Gamma^\prime$. By assumption, $\Gamma^\prime$ contains at least one vertex ${\bf x}^+ \in I^+$. 
But then $\hat{\bf T}^+ \cdot {\bf q}^+=-t{\bf e}_1\cdot {\bf e}_2=0$ which contradicts
(\ref{ns2}). This contradiction shows that the system (\ref{ns1}), (\ref{ns2})
has no solutions.



\begin{thebibliography}{}
\bibitem{Batch} \textsc{Batchelor, G. K. \& Green, J. T.} 1972 
The determination
of the bulk stress in a suspension of spherical particles to order
$c^2$. {\em J. Fluid Mech.} {\bf 56} Part 3,401--427.
\bibitem{Bach} \textsc{Bakhvalov, N. \& Panasenko, G.} 1989
{\em Homogenization: averaging processes in periodic media}. Kluwer.
\bibitem{BLP} \textsc{Bensoussan, A., Lions, J. L., \& Papanicolaou, G.} 1978 {\em Asymptotic analysis in periodic structures}.
North-Holland.
\bibitem{BBP} 
\textsc{Berlyand, L., Borcea, L \&  Panchenko, A.} 2003 Network approximation
for effective viscosity of concentrated suspensions with complex geometries.
{\em SIAM Journ. Math. Anal.}, to appear. 
\bibitem{BK}
\textsc{Berlyand, L. \&  Kolpakov, A.} 2001 Network approximation
in the limit of small inter-particle distance of the effective
properties of a high-contrast random dispersed composite.
{\em Arch. Rat. Mech. Anal.} {\bf 159}, 179--227.
\bibitem{Carr} 
\textsc{Carreau, P. J. \& Cotton, F.} 2002 Rheological properties
of concentrated suspensions. In: {\em Transport processes in bubbles, drops and particles}.
D. De Kee and R. P. Chhabra, eds., Taylor \& Fransis.
\bibitem{Cous}\textsc{Coussot, P.} 2002 Flows of concentrated
granular mixtures. In: {\em Transport processes in bubbles, drops and particles}.
D. De Kee adn R. P. Chhabra, eds., Taylor \& Fransis.
\bibitem{Einst}
\textsc{Einstein, A.} 1906 Eine neue Bestimmung der Molek\"{u}ldimensionen,
{\em Annln. Phys.} {\bf 19} p. 289 and {\bf 34} p. 591.
\bibitem{FA} \textsc{
Frankel, N. A. \& Akrivos, A.} 1967 On the Viscosity of
a Concentrated Suspension of Solid Spheres.  {\em Chemical Engineering Science}
{\bf 22}, 847--853.
\bibitem{JKO}
\textsc{Jikov, V., Kozlov, S. \& Oleinik, O.}
1994 {\em Homogenization of differential operators and integral functionals}.
Springer.
\bibitem{Keller} \textsc{Nunan. K. C. \&  Keller, J. B.} 1984 
Effective viscosity of periodic suspensions
{\em Journ. Fluid Mech.} {\bf 142}, 269--287. 
\bibitem{SP} \textsc{Sanchez-Palencia, E.} 1980
{\em Non-homogeneous media and vibration theory}. Springer.
\bibitem{Sh} \textsc{Schowalter, W. R.} 1978
{\em Mechanics of Non-Newtonian Fluids}. Pergamon Press.
\bibitem{Brady} \textsc{Sierou, A. \&  Brady, J. F.} 2001
Accelerated Stokesian
Dynamic simulations. {\em J. Fluid Mech.} {\bf 448}, 115--146.
\bibitem{ShP} \textsc{Shikata, T \& Pearson, D. S.} 1994
Viscoelastic behavior of 
concentrated spherical suspensions. {\em Journ. Rheology} {\bf 38}, 601--616.
\bibitem{Shook}
\textsc{Shook, C. A. \& Rocko, M. C.} 1991 {\em Slurry flow. Principles and practice.}
Butterworth-Heinemann.
\bibitem{VW} \textsc{Van der Werff, J. C., de Kruif, C. G.,
Blom, C,  \& Mellema, J.} 1989
Linear viscoelastic behavior of dense hard-sphere dispersions.
{\em Phys. Rev. A}. {\bf 39}, 795--807.
\end{thebibliography}
\end{document}